\documentclass[11pt,a4paper]{article}
\usepackage[utf8]{inputenc}
\usepackage{amsmath,amsthm,amssymb,amsfonts}
\usepackage{scalerel}
\usepackage{mathtools}
\usepackage{graphicx}
\usepackage{subfigure}
\usepackage{mathrsfs}
\usepackage{bbold}
\usepackage[]{units}
\usepackage{bm}
\usepackage{braket}
\usepackage{color}
\usepackage{makecell}
\usepackage{enumitem}
\usepackage{upgreek}
\usepackage{blindtext}
\usepackage{graphicx}
\usepackage{verbatim} 

\usepackage[dvipsnames]{xcolor}
\usepackage{bbm}
\usepackage[bb=boondox]{mathalfa}
\usepackage{array}
\usepackage{multirow}
\usepackage{tabularx}
\usepackage{float}
\usepackage{graphicx}
\usepackage{dcolumn}
\usepackage{bm}
\usepackage{amsmath,amsfonts,amssymb}
\usepackage{slashed}
\usepackage{braket,xcolor}
\usepackage{verbatim}
\usepackage{multirow}
\usepackage{amsfonts, mathtools,resizegather}
\usepackage[export]{adjustbox}

\usepackage{caption,jheppubmod}

\title{\boldmath Bulk reconstruction for anti-symmetric and symmetric gauge fields: $p$-forms and the graviton}

\author[a]{Budhaditya Bhattacharjee\footnote{Corresponding author.}}
\author[a]{, Jyotirmoy Mukherjee}

\affiliation[a]{Center For High Energy Physics, Indian Institute of Science\\ C V Raman Avenue, Malleshwaram, Bengaluru}
\emailAdd{budhadityab@iisc.ac.in}
\emailAdd{jyotirmoym@iisc.ac.in}

\abstract{We consider the HKLL bulk reconstruction procedure for $p$- form fields and graviton in empty AdS$_{d + 1}$. We derive spacelike bulk reconstruction kernels for the $p$-forms and the graviton in the Poincar\'e patch of $AdS_{d+1}$. The kernels are first derived via a mode-sum approach in arbitrary even dimensions. The appropriate AdS-covariant fields are identified and corresponding kernels obtained via the mode sum approach for these fields. We present arguments for casting these kernels in terms of the AdS chordal distance. Introducing an antipodal-like mapping, the kernels are cast in a spacelike form. An alternative derivation is presented for these kernels, by using a chordal Green’s function approach. From the asymptotic expansion of the bulk fields and using Green’s theorem, we determine the spacelike kernels for both modes of $p$-forms and the graviton, which are shown to agree with the kernels obtained by the mode-sum approach. The massive $p-$ form fields, as well as the relevant Brietenlohner-Freedman bounds are also discussed.} 

\begin{document} 
\maketitle
\flushbottom

\section{Introduction}
\label{sec:intro}
The AdS/CFT correspondence \cite{Witten:1998qj, Gubser:1998bc} is a duality between a theory of quantum gravity in Anti-de Sitter spacetime and a conformal field theory living on a codimension$-1$ surface. It is a well-studied instance of the holographic principle \cite{Maldacena:1997re}. One of the primary questions is to describe bulk physics in terms of holographic boundary operators. One way to proceed with this problem is to write bulk fields that solve the bulk wave equations in terms of the boundary operators. Semi-classically, one of the ways to achieve the same is by inverting the extrapolate AdS/CFT dictionary. This is described in \cite{Banks:1998dd, Bena:1999jv}. Explicit calculations were done in the seminal works of Hamilton, Kabat, Lifshitz and Lowe \cite{Hamilton:2005ju, Hamilton:2006az, Hamilton:2006fh, Hamilton:2007wj}, with further extensions covered in \cite{Kabat:2012hp,Sarkar:2014dma,Sarkar:2014jia}. The general idea and some of the explicit calculations were reviewed in \cite{DeJonckheere:2017qkk, Kajuri:2020vxf}.

The HKLL reconstruction is generally developed for the normalizable mode \cite{Hamilton:2006az}. This is a natural thing to do since only the normalizable mode has a well-defined holographic dual in terms of CFT operators. Instead, the non-normalizable mode is best thought of as deformations of the CFT. However, from the bulk point of view, there is no a-priori reason preventing a formal construction of the HKLL kind for the non-normalizable mode. An additional motivation lies in the fact that within the Breitenlohner-Freedman mass window \cite{Breitenlohner:1982bm}, both modes are acceptable. To be explicit, it is well-known BF bound for $p$-forms on $AdS_{d+1}$ is  $m^2_{\rm{BF}}\geq -(\frac{d}{2}-p)^2$ \cite{Witten:1998qj}.  Note that when BF bound is saturated, normalizable and the non-normalizable solutions become identical. Therefore, both solutions of
the bulk wave equation of $p$-forms are acceptable as genuine operators in the CFT spectrum. So it is required to develop the non-normalizable modes to get the complete picture in CFT. A similar statement holds true for graviton and symmetric higher spin fields. In \cite{Bhattacharjee:2022ehq}, the formal kernels of scalar were derived for both modes in Poincar\'e and Global coordinates. The discussions there cover some technical aspects of the prescription to evaluate the kernels via the mode-sum and Green’s function approaches.

From \cite{Bhattacharjee:2022ehq} and \cite{Heemskerk:2012np, Foit:2019nsr, Sarkar:2014dma, Kabat:2012hp}, a natural question one can ask is to consider the reconstruction prescription for the fields with gauge symmetries. Purely from the perspective of solutions of differential equations, it is an interesting direction to pursue due to the rich technical structure. We expect to learn about the organization of solutions of such differential equations. In this work, we evaluate the kernels in Poincar\'e coordinates for fields with spin, focusing explicitly on the $p$-forms and the graviton. We derive kernels in the Poincar\'e patch for $p$-forms and gravitons for normalizable as well as for non-normalizable modes. The construction is complicated by the fact that the gauge needs to be fixed and the bulk field operators expressed in terms of gauge invariant boundary currents.
\begin{align}
    A^p(b)&=\int \mathbf{K}_N(b;x) j^p_{N}(x)+\int \mathbf{K}_{nN}(b;x)j^p_{nN}(x),
\end{align}
where $A^p$ denotes contravariant antisymmetric $p$-forms and  $j^p$ is the boundary current. Here b stands for the location of the bulk field. The nature of organization of the solutions of bulk equation of motion has been studied in depth \cite{Balasubramanian:1998sn, Bhattacharjee:2019xhb}.In order to obtain the bulk field, one requires to integrate over the boundary coordinate $x$. Since the bulk wave equation is a second-order differential equation, it is quite natural to have two independent solutions, which we call normalizable and non-normalizable modes. Our goal is to isolate the kernels for these modes in $p$-forms and the graviton.

The organization of the paper is as follows. In section \eqref{review}, we briefly review  HKLL reconstruction of the massive scalar field. We also present kernels for the two modes in the massless limit that will be used as a cross-check for $p$-forms at $p\rightarrow 0$ limit.   We derive kernels in Poincar\'e patch of AdS for both modes in two ways – using a mode sum approach as well as
a spacelike Green’s function approach. In section \eqref{msum}, we obtain kernels for both modes for $p$-forms. We first fix the gauge of the bulk fields using a holographic gauge fixing procedure \cite{Sarkar:2014dma} and obtain bulk equations for $p$-forms. Using the mode sum approach, we obtain kernels for both modes in even and odd AdS. In section \eqref{g1}, we derive kernels for both modes of the graviton using the mode sum approach in the Poincar\'e patch of AdS in even and odd dimensions. Using some transformation properties of the hypergeometric function, we extract the covariant piece from the kernels for both modes. In section \eqref{sec5}, we present a direct method for constructing the spacelike kernel for $p$-forms and the graviton. We show that under suitable scaling of the bulk fields, bulk wave equation can be written as AdS covariant form. From the asymptotic expansion of the rescaled bulk fields and using Green’s theorem, we read off the spacelike kernels for both modes of $p$-forms and the graviton. These kernels precisely match the kernels obtained in the mode-sum approach.

\section{Review of HKLL reconstruction}
\label{review}
The Lorentzian AdS/CFT dictionary tells us that local operators in the bulk geometry can be described in terms of non-local operators in the boundary CFT \cite{Gubser:1998bc, Balasubramanian:1998de, Balasubramanian:1998sn, Witten:1998qj, Banks:1998dd, Klebanov:1999tb}. The explicit procedure for establishing this map is known as Bulk Reconstruction. In this work, we focus on the procedure developed by Hamilton, Kabat, Lifschytz and Lowe. The HKLL\footnote{for the authors' Hamilton, Kabat, Lifschytz and Lowe} bulk reconstruction scheme was originally developed in the series of papers \cite{Hamilton:2005ju,Hamilton:2006az,Hamilton:2006fh,Hamilton:2007wj,Kabat:2012hp}. Further work in this direction has since been carried out \cite{Foit:2019nsr,Terashima:2017gmc,Terashima:2020uqu,Terashima:2021klf, Sarkar:2014dma,Sarkar:2014jia, Kajuri:2018wow,Kajuri:2019kmr,Kajuri:2020bvi, Dey:2021vke, Aoki:2023lgr, Bhattacharjee:2022ehq}. For a detailed review of HKLL bulk reconstruction, see \cite{DeJonckheere:2017qkk,Kajuri:2020vxf}. 

The HKLL bulk reconstruction procedure is executed at the semiclassical level, where large$-N$ and large-t'Hooft coupling is assumed. It is then reasonable to treat bulk fields (or operators) as free. At the level of a mathematical expression, the bulk reconstruction map is represented as follows
\begin{align}
    \phi(z, x) = \int_{\partial \mathcal{M}}\mathbf{K}(z,x; x')\mathcal{O}(x')\mathrm{d}^{d}x'\label{blkrec}
\end{align}
where $\phi$ stands for the bulk operator and $\mathcal{O}(x)$ is the corresponding CFT operator on the boundary. The integration is over the boundary surface $\partial\mathcal{M}$, and $z$ is the holographic direction. In the original HKLL papers, the bulk field was chosen to have a specific fall-off behavior at the boundary ($z \rightarrow 0$) given by $\phi(z, x) \sim z^{\Delta}\phi_0(x)$. Such fields are known as \textit{normalizable} when $\Delta > \frac{d}{2} - 1$ (the Brietenlohner-Freedman bound \cite{Breitenlohner:1982bm}), which follows from unitarity of the boundary CFT. The exponent $\Delta$ also corresponds to the conformal dimension of the CFT operator $\mathcal{O}(x)$. 

The HKLL reconstruction provides us with a prescription to explicitly evaluate the \textit{kernel} $K(z, x; x')$ in \eqref{blkrec}. The idea is to solve the bulk wave equation for the free field and then invert the equations by choosing the appropriate boundary value of the bulk field. In the next sub-section, we review some results for the kernel for the free scalar field.
\subsection{HKLL reconstruction for scalar}
\label{sec:review}
In this section, we briefly review the main results of \cite{Hamilton:2006az, Bhattacharjee:2022ehq}, where the bulk reconstruction kernels were explicitly evaluated for the free scalar field. In \cite{Bhattacharjee:2022ehq}, the \textit{non-normalizable mode} was also considered aside from the usual normalizable mode. Additionally, in \cite{Aoki:2023lgr, DelGrosso:2019gow}, the kernel for the normalizable mode was analytically continued beyond the Brietenlohner-Freedman (BF) bound. One of the main motivations for studying the non-normalizable mode lies in the fact that the two modes are indistinguishable (by their fall-off behaviors) within the BF \textit{window} $\frac{d}{2} - 1 \leq \Delta \leq \frac{d}{2} + 1$. Therefore both modes are identified with CFT operators on the boundary. 

The HKLL bulk reconstruction procedure begins with solving the bulk wave equation (in empty AdS$_{d + 1}$) for the free massive scalar. It is written (in Lorentzian Poincar\'e coordinates $\{z, t, x\}$) as follows
\begin{align}
    \left(z^2\partial^2_{z} + z^2\partial^2_{x} - z(d - 1)\partial_{z} - m^2\right)\Phi(z,\vec{x}) = 0
\end{align}
where $\{x\} \equiv \{t, \vec{x}\}$. The solution to this equation is
\begin{align}
    \Phi(z,\vec{x}) = \int \frac{d^{d}q}{(2\pi)^{d}}e^{i q. x}z^{\frac{d}{2}}\left(a(q)J_{\nu}(q z) + b(q)Y_{\nu}(q z)\right)
\end{align}
with $\nu = \sqrt{\frac{d^2}{4} + m^2} \equiv \Delta - \frac{d}{2}$. To invert this solution and determine the two undefined functions $a(q)$ and $b(q)$, the boundary data needs to be fixed. The natural choice for the free massive scalar field is
\begin{align}
    \Phi(z, x)\vert_{z \rightarrow 0} \sim z^{\Delta}\phi_{0}(x) + z^{d - \Delta}j_{0}(x)\label{blkField1}
\end{align}
where the two pieces of boundary data $\phi_{0}$ and $j_{0}$ correspond to the normalisable and non-normalisable modes respectively. Thus, the bulk fields can be written as
\begin{align}
    \Phi(z, x) = \int \mathbf{K}_{N}(z, x; x')\phi_{0}(x')\mathrm{d}^{d}x' + \int \mathbf{K}_{nN}(z, x; x')j_{0}(x')\mathrm{d}^{d}x'\label{blkField2}
\end{align}
By comparing \eqref{blkField1} and \eqref{blkField2} (and after some algebra), the two kernels $\mathbf{K}_{N}$ and $\mathbf{K}_{n N}$ can be evaluated. The expressions \cite{Bhattacharjee:2022ehq} are as follows
\begin{align}
    \mathbf{K}_{N}(z, x; x') &= \frac{\Gamma(d/2)}{2 \pi^{d/2}}\frac{z^{\Delta}}{(\sqrt{\Delta \vec{x}^2 - \Delta t^2})^{d}}\;_{2}F_{1}\left(\frac{d}{2}, 1; 1 - \frac{d}{2} + \Delta; - \frac{z^2}{\Delta \vec{x}^2 - \Delta t^2} \right) \label{1KerN} \\
    \mathbf{K}_{n N}(z, x; x') &= \frac{\Gamma(d/2)}{2 \pi^{d/2}}\frac{z^{d - \Delta}}{(\sqrt{\Delta \vec{x}^2 - \Delta t^2})^{d}}\;_{2}F_{1}\left(\frac{d}{2}, 1; 1 + \frac{d}{2} - \Delta; - \frac{z^2}{\Delta \vec{x}^2 - \Delta t^2} \right) \label{1KernN}
\end{align}
where $\Delta x^2 = \vert \vec{x} - \vec{x}' \vert^2$ and $\Delta t^2 = (t - t')^2$. It is important to note that the kernels written here are not explicitly AdS covariant. However, directly checking the action of AdS isometries proves that the fields are AdS covariant \cite{Hamilton:2006az}. Additionally, the support of the kernel in \eqref{blkField2} is over the full boundary, whereas the causally connected region of the bulk point $z, x$ only connects it to the spacelike separated points. So ideally, the kernel should be both spacelike and AdS-covariant.

This is the mode-sum approach to evaluating the bulk reconstruction kernels for the two modes. An equivalent approach is via using a spacelike Green's function. In \cite{Hamilton:2006az}, a Green's function is constructed for the normalizable mode, and a spacelike, AdS-covariant kernel is obtained via Green's theorem. In a similar spirit, a Green's function is considered for the non-normalizable mode \cite{Bhattacharjee:2022ehq}, and a spacelike, AdS-covariant kernel is obtained. The respective kernels are as follows
\begin{align}
    \mathbf{K}_{N} &= \frac{(-1)^{\frac{d-1}{2}}2^{\Delta - d}\Gamma(1 + \Delta - \frac{d}{2})}{2\pi^{\frac{d}{2}}\Gamma(1 + \Delta - d)}\lim_{z' \rightarrow 0}(\sigma z')^{\Delta - d}\theta(\text{spacelike}) \label{2kerN} \\
    \mathbf{K}_{n N} &= -\frac{2^{-\Delta}\Gamma(\Delta)\tan\pi\Delta}{2\pi^{d/2}\Gamma(\Delta - \frac{d}{2})}\lim_{z' \rightarrow 0}(\sigma z')^{-\Delta}\theta(\text{spacelike})\label{2kernN}
\end{align}
where $\sigma$ is the (covariant) geodesic length in pure AdS. It is clear that \eqref{1KerN} and \eqref{1KernN} are not equal to \eqref{2kerN} and \eqref{2kernN} respectively. However, it is important to note that the kernel is not unique, and it is possible to add or subtract terms to it as long as those terms vanish when integrated against the boundary field(s). Using this and an antipodal map-inspired redefinition of the boundary data \cite{Bhattacharjee:2022ehq}, it can be shown that \eqref{1KerN} and \eqref{1KernN} are equivalent to \eqref{2kerN} and \eqref{2kernN} respectively.

The case of the massless free scalar field is of special importance to this work. For such a field, the kernels are given by
\begin{align}
    \mathbf{K}_{N} &= \frac{(-1)^{\frac{d-1}{2}}\Gamma(1 + \frac{d}{2})}{2\pi^{\frac{d}{2}}}\theta(\text{spacelike}) \label{2kerNm0} \\
    \mathbf{K}_{n N} &= -\frac{2^{-d}\Gamma(d)\tan\pi d}{2\pi^{d/2}\Gamma(\frac{d}{2})}\lim_{z' \rightarrow 0}(\sigma z')^{-d}\theta(\text{spacelike})\label{2kernNm0}
\end{align}
Likewise, the expressions \eqref{1KerN} and \eqref{1KernN} become
\begin{align}
    \mathbf{K}_{N}(z, x; x') &= \frac{\Gamma(d/2)}{2 \pi^{d/2}}\frac{z^{d}}{(\sqrt{\Delta x^2 - \Delta t^2})^{d}}\;_{2}F_{1}\left(\frac{d}{2}, 1; 1 + \frac{d}{2}; - \frac{z^2}{\Delta x^2 - \Delta t^2} \right) \label{1KerNm0} \\
    \mathbf{K}_{n N}(z, x; x') &= \frac{\Gamma(d/2)}{2 \pi^{d/2}}\frac{1}{(\sqrt{\Delta x^2 - \Delta t^2})^{d}}\;_{2}F_{1}\left(\frac{d}{2}, 1; 1 - \frac{d}{2}; - \frac{z^2}{\Delta x^2 - \Delta t^2} \right) \label{1KernNm0}
\end{align}
As a consistency check for our results of the $p$-form field, we consider the massless free scalar field, which is identical to the $p = 0$ form field. 

With this, we conclude a brief review of the HKLL bulk reconstruction procedure for free massive scalar fields in empty AdS. For more details and technicalities, the reader is directed to the references mentioned at the beginning of this section. In the following sections, we extend this procedure to the free massless $p$-form field and the linearized graviton in empty AdS. 
\section{HKLL reconstruction for $p$-forms}\label{msum}

Theories with gauge symmetries do not admit local
gauge-invariant observables. One can construct gauge-invariant operators using Wilson loops or the product of Wilson loops. 
\begin{align}
    W(z,x)=\lim_{\epsilon\rightarrow 0}\oint_C A,
\end{align}
where $A$ is $p$-form gauge potential. Therefore, one may wonder if the constructed bulk field operators have any physical meaning, in particular in a gravitational theory where there are no local diffeomorphism invariant observables. But it has been shown that non-local gauge invariant operators can be traded for local
operators in a fixed gauge \cite{Heemskerk:2012np}. However, commutators of these operators can exhibit non-locality due to the Gauss' law constraint.

In this paper, we follow a similar prescription to obtain the smearing functions, and  these operators in the bulk of AdS can be mapped to operators in a boundary
theory through the extrapolate dictionary\cite{Polchinski:2010hw}
\begin{align}
    \lim_{z\rightarrow 0}\sqrt{-g}F^{z,\mu_1\cdots \mu_p}(z,x)=j^{\mu_1\cdots \mu_p}(x).
\end{align}
In this section, we explicitly derive the spacelike bulk reconstruction kernel
corresponding to the two modes of $p$-forms. We focus our attention on the Poincar\'e patch of AdS. To derive the bulk reconstruction kernel, the first step is to write the bulk wave equation of $p$-form fields on the Poincar\'e patch.
The equation of motion of $p$-forms is given by
\begin{align}
    \nabla_{M}F^{M,J_{1},\dots,J_{p}} = 0,\label{maxe}
\end{align}
where $F^{M,J_{1},\dots,J_{p}}$ is the field strength corresponding to the antisymmetric gauge potential $ A^{J_{1},\dots,J_{p}}$.  We use capitalized Latin indices $A, B,\dots$ to indicate bulk coordinates, while Greek indices $\alpha,\beta,\dots$ indicate boundary coordinates. As it is known that the action has $U(1)$ gauge symmetry, we make certain gauge choices to obtain the bulk equation corresponding to the transverse gauge potential. We choose
\begin{align}
    A^{J_{1},\dots,J_{p-1},z} &= 0 
\end{align}
 as a part of our gauge choice and use $\nabla_{\alpha_1}A^{\alpha_1,\alpha_2,\cdots,\alpha_p}=0$ to derive the equation of motion. We impose the gauge conditions in \eqref{maxe} and write the bulk wave equation of $p$-forms on the Poincar\'e patch. \footnote{Details of this derivation are shown in the Appendix \eqref{appA}}
\begin{align}
     z^{2}\partial^2_{z}A_{J} + (2 p - d + 1)z\partial_{z}A_{J} + z^2 \partial_{\alpha}\partial^{\alpha}A_{J} = 0,\quad\quad\quad A_{J_{1},\dots,J_{p}} \equiv A_{J}
\end{align}
The solution to the wave equation in this background is obtained as
\begin{align}
      A_{J}(z, x) = \int_{q \geq 0}\frac{\mathrm{d}^{d}q}{(2 \pi)^{d}}z^{\frac{1}{2} (d-2 p)}\left(c_{J}(q) J_{\frac{1}{2} (d-2 p)}(|q| z)+ d_{J}(q) Y_{\frac{1}{2} (d-2 p)}(|q| z)\right)e^{i q. x}.
\end{align}
Here $x\equiv (t,\Vec{x})$ and $q \equiv (\omega, \Vec{k})$ with $q=\sqrt{\omega^2-|k|^2}.$ Let us also define a parameter $\nu=\frac{d}{2}-p$. The near boundary behavior of the solution can be extracted from the expansion of the Bessel functions at $z=0$. For now, we focus on the case where $\frac{d}{2}-p=\nu\neq \text{Integer}$. This corresponds to the case where $d$ is odd, i.e., Even AdS. The \textit{massive} $p$-form is also briefly discussed in Appendix.\ref{AppMassP}.

\subsection{Even AdS}
We now focus on the mode
expansions for $p$-forms and write down the kernels in even AdS$_{d + 1}$. The odd AdS case is discussed in the Appendix \ref{appB}. 

In even dimensional AdS, there are two independent solutions to the bulk wave equation, which are $J_{\nu}(|q|z)$ and $J_{-\nu}(|q|z)$. This is due to the fact that $\nu \in \frac{1}{2}\mathbb{Z}$ for odd $d$ and therefore
\begin{align}\label{brel}
Y_{\nu}(|q|z)=\frac{J_{\nu}(|q|z)\cos(\nu\pi)-J_{-\nu}(|q|z)}{\sin(\nu\pi)}.
\end{align}
The solution to the wave equation in this background can be expressed
\begin{align}
      A_{J}(z, x) = \int_{q \geq 0}\frac{\mathrm{d}^{d}q}{(2 \pi)^{d}}z^{\nu}\Big[a_{J,\nu}(q) J_{\nu}(|q| z)+ b_{J,\nu}(q) J_{-\nu}(|q| z)\Big]e^{i q. x}.
\end{align}
The coefficients $a_{J,\nu}$ and $b_{J,\nu}$ are given by
\begin{align}
    a_{J,\nu}(|q|)=c_{J}(q)+\cot(|q|\nu\pi) d_{J}(|q|),\quad\quad b_{J,\nu}(|q|)=-\frac{d_J(|q|)}{\sin(\nu\pi)}.
\end{align}
The coefficient of each term in the expansion of the solution around $z=0$ can be expressed as a function of x.
\begin{align}
    A_{J,\nu}(x,z)&=z^{-p}\sum_{n=0}^{\infty}\left(z^{\Delta+2n}\phi_{J}^{(2 n)}(x)+z^{d-\Delta+2n}j_{J}^{(2 n)}(x)
    \right)\nonumber\\
    &=  A^{N}_{J}(z, x)+ A^{n N}_{J}(z, x) , 
\end{align}
where $\Delta=d-p=\nu+\frac{d}{2}$ and two modes are given by
\begin{align}
    A^{N}_{J}(z, x) = \sum_{n = 0}^{\infty} z^{2 n+2\nu}\phi_{J}^{(2 n)}(x) ,\quad\quad
    A^{n N}_{J}(z, x) = \sum_{n = 0}^{\infty} z^{2 n }j_{J}^{(2 n)}(x) \label{expansionPform}
\end{align}
The coefficients in each order can be expressed
\begin{align}\label{a1}
    \phi_{J}^{(2n)}(x) &= \frac{(-1)^{n}}{2^{\nu}4^n\Gamma( n + 1)\Gamma(\nu + n + 1)}\int_{q \geq 0}\frac{\mathrm{d}^{d} q}{(2 \pi)^{d}}a_{J,\nu}(q)|q|^{2n+\nu}e^{i q. x} \\
    j_{J}^{(2n)}(x)& = \frac{(-1)^{n}}{2^{-\nu}4^n\Gamma( n + 1)\Gamma(-\nu + n + 1)}\int_{q \geq 0}\frac{\mathrm{d}^{d} q}{(2 \pi)^{d}}b_{J,\nu}(q)|q|^{2n-\nu}e^{i q. x}\label{a2}
\end{align}
\subsection*{Kernels as mode-sum integrals}
The integrals in \eqref{a1} and \eqref{a2} can be inverted to obtain $a_{J,\nu}(|q|)$ and $b_{J,\nu}(|q|)$. In order to do that, we pick two independent pieces of data from the expansion coefficients of the two modes, i.e., $\phi_{J,2n}(x)$ and $j_{J,2n}(x)$. We choose coefficients corresponding to $n=0$  as data, and once these data are extracted, then the rest of the terms can be evaluated in terms of them. The resultant expressions for $a_{J,\nu}(|q|)$ and $b_{J,\nu}(|q|)$ are obtained as Fourier transforms of the boundary fields.

The coefficients corresponding to $n=0$ of the mode expansions are given by
\begin{align}\label{kerp}
    \phi_{J}^{(0)}(x) &= \frac{1}{2^{\nu}\Gamma(\nu  + 1)}\int_{q \geq 0}\frac{\mathrm{d}^{d} q}{(2 \pi)^{d}}a_{J,\nu}(q)|q|^{\nu}e^{i q. x}\\
    j_{J}^{(0)}(x)& = \frac{1}{2^{-\nu}\Gamma(-\nu  + 1)}\int_{q \geq 0}\frac{\mathrm{d}^{d} q}{(2 \pi)^{d}}b_{J,\nu}(q)|q|^{-\nu}e^{i q. x}
\end{align}
Using Fourier transformation, we extract $a_{J}(q)$ and $b_{J}(q)$ in terms of these boundary data and $q$.
\begin{align}
    a_{J,\nu}(|q|) &= \Gamma(\nu+1)2^{\nu}\int |q|^{-\nu}\phi_{J}^{( 0)}(x')e^{-i q. x'}\mathrm{d}^{d}x' \\
    b_{J,\nu}(|q|) &=\frac{\Gamma(1-\nu)}{2^{\nu}}\int |q|^{\nu}j_{J}^{( 0)}(x')e^{- i q. x'}\mathrm{d}^{d}x'
\end{align}
From these expressions, it is easy to read off the kernel integrals
\begin{align}
    \mathbf{K}_{N}(z,x;x') =  2^{\nu}\Gamma(\nu+1)\int_{q \geq 0} \frac{\mathrm{d}^{d}q}{(2 \pi)^{d}} z^{\nu}|q|^{-\nu}J_{\nu}(|q| z)e^{i q. (x - x')} \label{NormEvenAdSkernel} \\
    \mathbf{K}_{nN}(z,x;x') =  2^{-\nu}\Gamma(1-\nu)\int_{q \geq 0} \frac{\mathrm{d}^{d}q}{(2 \pi)^{d}} z^{\nu}|q|^{\nu}J_{-\nu}(|q| z)e^{i q. (x - x')} \label{nonNormEvenAdSkernel}
\end{align}
\subsection*{Explicit evaluation of the Poincar\'e kernel integrals}
In this section, we evaluate the kernels \eqref{NormEvenAdSkernel} and \eqref{nonNormEvenAdSkernel} in the Poincar\'e patch of the even-dimensional AdS space.  We use the following integral \cite{Bhattacharjee:2022ehq}, where $\zeta$ stands for any Bessel function and $\mu, \nu$ are real numbers. 
\begin{align}
    \int_{|q| \geq 0}\frac{\mathrm{d}^{d}q}{(2\pi)^{d}}|q|^{\mu}\zeta_{\nu}(q z)e^{i q. (x - x')} = \frac{1}{\pi(2\pi)^{d/2}X^{d/2 - 1}}\int_{x = 0}^{\infty}x^{\mu + d/2}\zeta_{\nu}(x z)K_{\frac{d}{2}-1}(x X)\mathrm{d}x \label{identity_integral-0}
\end{align}
For the normalizable kernel, we find that in terms of the notation in the above expression, we have $\zeta_{\nu} = J_{\nu}$ and $\mu = -\nu$. Thus the integral becomes
\begin{align}
    \mathbf{K}_{N}(z, x; x') =  \frac{2^{\nu}\Gamma(\nu+1)z^{\nu}}{\pi (2 \pi)^{d/2} X^{d/2 - 1}}\int_{ 0}^{\infty}\mathrm{d}x x^{\frac{d}{2}-\nu}J_{\nu}(x z)K_{\frac{d}{2} - 1}(x X)
\end{align}
To evaluate this integral, we need to use the following identity
\begin{align}
    \int_{0}^{\infty}x^{-\lambda}K_{\mu}(a x)J_{\nu}(b x) &= \frac{b^{\nu}\Gamma(\frac{\nu -\lambda + \mu + 1}{2})\Gamma(\frac{\nu -\lambda - \mu + 1}{2})}{2^{\lambda + 1}\Gamma(\nu + 1)a^{-\lambda + \nu + 1}}\notag\\
    &\times\;_{2}F_{1}\left(\frac{\nu -\lambda + \mu + 1}{2}, \frac{\nu -\lambda - \mu + 1}{2}; \nu + 1; -\frac{b^2}{a^2}\right)\label{identity_integral-1}
\end{align}
Identifying $\lambda = \nu-\frac{d}{2}=-p$ and $\mu = \frac{d}{2} - 1 $(with $a = X$ and $b = z$), we have the following expression of the kernel corresponding to the normalizable modes
\begin{align}
    \mathbf{K}_{N}(z, x; x')
    = \frac{z^{\Delta -p} \Gamma(d/2)}{2 \pi^{\frac{d}{2} + 1}X^{d}}\;_{2}F_{1}\left(\frac{d}{2}, 1;1-\frac{d}{2}+\Delta; - \frac{z^2}{X^2}\right)\label{EvenAdSkernelp1}
\end{align}
For $p=0$ or the $0$-form, the kernel reduces to the kernel of a massless scalar. Similarly, for $p = 1$ the kernel reduces to that of a free Maxwell field in even AdS
\begin{align}
     \mathbf{K}_{N}(z, x; x')|_{p=0} &= \frac{z^{\Delta} \Gamma(d/2)}{2 \pi^{\frac{d}{2} + 1}X^{d}}\;_{2}F_{1}\left(\frac{d}{2}, 1;\frac{d}{2}+1; - \frac{z^2}{X^2}\right)\label{p0} \\
     \mathbf{K}_{N}(z, x; x')|_{p=1} &= \frac{z^{\Delta-1} \Gamma(d/2)}{2 \pi^{\frac{d}{2} + 1}X^{d}}\;_{2}F_{1}\left(\frac{d}{2}, 1;\frac{d}{2}; - \frac{z^2}{X^2}\right)
\end{align}
The kernel in \eqref{p0} provides a consistency check of our computations. Similarly, the kernel for the non-normalizable mode evaluates to
\begin{align}
    \mathbf{K}_{nN}(z, x; x') = \frac{\Gamma(d/2)}{2 \pi^{d/2 + 1}X^{d}}\;_{2}F_{1}\left(\frac{d}{2}, 1; 1 - \Delta + \frac{d}{2}; - \frac{z^2}{X^2}\right)\label{EvenAdSNNkernelp1}
\end{align}

\subsection{Covariant form of the kernels}
Here we consider certain transformations that cast the kernels  \eqref{EvenAdSkernelp1} and \eqref{EvenAdSNNkernelp1} in terms of an AdS-covariant piece (and an extra term). 

Using the following variable transformation identity of the hypergeometric $\;_{2}F_{1}$
\begin{align}
    \;_{2}F_{1}(a, b; c; z) = &\frac{\Gamma(c)\Gamma(b - a)}{\Gamma(b)\Gamma(c-a)}(-z)^{-a}\;_{2}F_{1}\left(a, 1 - c + a; 1 - b + a; \frac{1}{z}\right) \notag\\ & + \frac{\Gamma(c)\Gamma(a - b)}{\Gamma(a)\Gamma(c-b)}(-z)^{-b}\;_{2}F_{1}\left(b, 1 - c + b; 1 - a + b; \frac{1}{z}\right)\label{hypIden}
\end{align}
we rewrite \eqref{EvenAdSkernelp1} as follows
\begin{align}
    z^{p}\mathbf{K}_{N} &= \frac{\Gamma(\frac{d}{2} - 1)\Gamma(1 + \frac{d}{2} - p)}{2\pi^{\frac{d}{2} + 1}\Gamma(\frac{d}{2} - p)}\frac{z^{d - p - 2}}{X^{d - 2}}\;_{2}F_{1}\left(1, 1 - \frac{d}{2} + p; 2 - \frac{d}{2}; - \frac{X^2}{z^2}\right) \notag\\ &+ \frac{(-1)^{d/2}\Gamma(1 + \frac{d}{2} - p)2^{\Delta - d}}{2\pi^{\frac{d}{2}}\Gamma(1 - p)}\lim_{z'\rightarrow 0}(\sigma z')^{-p}\label{pformAdSCovKerN}
\end{align}
and similarly \eqref{EvenAdSNNkernelp1} is written as
\begin{align}
    z^{p}\mathbf{K}_{n N} &= \frac{\Gamma(\frac{d}{2} - 1)\Gamma(1 - \frac{d}{2} + p)z^{p-2}}{2\pi^{\frac{d}{2} + 1}X^{d-2}}\;_{2}F_{1}\left(1, 1 + \frac{d}{2} - p; 2 - \frac{d}{2}; - \frac{X^2}{z^2}\right) \notag \\ &+ \frac{(-1)^{\frac{d}{2}}\Gamma(1  - \frac{d}{2} + p)2^{p - d}}{\Gamma(1 - d + p)2\pi^{\frac{d}{2}}}\lim_{z' \rightarrow 0}(\sigma z')^{p - d}\label{pformAdSCovKernN}
\end{align}
Therefore, we note that the AdS-covariant piece in the respective kernels are the second terms in \eqref{pformAdSCovKerN} and \eqref{pformAdSCovKernN}. It is straightforward to see that there are no common terms between the first and second pieces of either of the above two expressions.

The first terms in \eqref{pformAdSCovKerN} and \eqref{pformAdSCovKernN} only have terms proportional to $z^{d - p - 2 - 2 n}$ and $z^{p - 2 - 2n}$ respectively. Such terms do not arise in the respective modes in \eqref{expansionPform}. This implies that these terms shall vanish when integrated against the corresponding boundary field. 

A plausible prescription for dropping these extra terms was discussed in Appendix D of \cite{Bhattacharjee:2022ehq}. In this case, the same can be employed by noting that there is a pole/branch-point at $X = 0$ in the first terms of both \eqref{pformAdSCovKerN} and \eqref{pformAdSCovKernN} (since $d$ is odd). {\color{black} Here we briefly recall the argument. It is similar in spirit to the arguments presented for~\cite{Hamilton:2006az,Hamilton:2005ju} for demonstrating that similar terms vanish. The argument relies on introducing a regulator $\pm i\epsilon$ when performing the integral $\int \mathbf{K}\phi$ against the boundary field $\phi$ in such a way that the only surviving contribution is from the AdS-covariant piece. Let us begin by considering the integral against positive energy modes, which leads to integrals of the form\footnote{Here $\lambda$ is the exponent, which gives rise to a pole/branch-point wherever the distance term vanishes. The exact value is not important.} (focusing on AdS$_4$)
\begin{align}
    I_A &= \int_{T,X,Y = -\infty}^{\infty}(T^2 - X^2 - Y^2)^{\lambda}e^{-i\omega t}e^{-i k_x X}e^{-i k_y Y}\mathrm{d}X \mathrm{d}Y \mathrm{d}T \\
    I_B &= \int_{T,X,Y = -\infty}^{\infty}(T^2 - X^2 - Y^2 - Z^2)^{\lambda}e^{-i\omega t}e^{-i k_x X}e^{-i k_y Y}\mathrm{d}X \mathrm{d}Y \mathrm{d}T
\end{align}
This is under the condition that $\omega > \vert k \vert$, due the boundary field being a positive energy mode. The next step is to introduce two different parameterizations of the coordinates for $Y > 0$ and $Y < 0$. For $Y > 0$, we set the parametrization as $X = R \cos\theta, Y  = R\sin\theta$ where $\theta$ is with respect to the positive $X$ axis and $R \in (0,\infty)$. For $Y < 0$, use the alternate parameterization $X = R'\cos\theta', Y = R'\sin\theta'$ where $\theta'$ is now with respect to the $-$ve $X$ axis, and thus $R' \in (-\infty,0)$. Note that both $\theta,\theta' \in (0,\pi)$ This is crucial to the argument, since it allows us to write the integrals as 
\begin{align}
    I_A &= \int_{T,R = -\infty}^{\infty}\int_{\theta = 0}^{\pi}R(T^2 - R^2)^\lambda e^{-i\omega T}e^{i k R \cos(\theta - \alpha)}\mathrm{d}R\mathrm{d}T\mathrm{d}\theta\\
    I_B &=  \int_{T,R = -\infty}^{\infty}\int_{\theta = 0}^{\pi}R(T^2 - R^2 - Z^2)^\lambda e^{-i\omega T}e^{i k R \cos(\theta - \alpha)}\mathrm{d}R\mathrm{d}T\mathrm{d}\theta
\end{align}
where we have use the parametrization $k_{x} = k\cos\alpha, k_y = k\sin\alpha$. Note that both the integrals have $R\in(-\infty,\infty)$ due to the choice of two different parametrizations. This is generalizable to higher dimensions, by choosing two different angles on the sphere with respect to the North and South poles in the two cases. Finally, using $U = T - R$ and $V =  T + R$ along with the modes $\omega_\pm = \omega \pm k\cos(\theta-\alpha) > 0$, we obtain
\begin{align}
    I_A &= \int_{U,V = -\infty}^\infty \left(\frac{U - V}{2}\right)(UV)^\lambda e^{-i\omega_{-}U - i \omega_{+}V}\mathrm{d}U\mathrm{d}V \\
    I_B &= \int_{U,V = -\infty}^\infty \left(\frac{U - V}{2}\right)(UV - Z^2)^\lambda e^{-i\omega_{-}U - i \omega_{+}V}\mathrm{d}U\mathrm{d}V
\end{align}
Note that the integral contour has to be closed in the LHP, since $\omega_\pm > 0$. Introducing a regulator $R \rightarrow R \pm i \epsilon$. This leads to the condition for the pole/branch-point in $I_A$ being
\begin{align}
    (U \mp i\epsilon)(V \pm i\epsilon) = 0
\end{align}
Say if we choose the regulator to be $+ i\epsilon$, then the pole for $U$ is at $U_0 = i\epsilon$. This lies outside the integral contour and thus $I_A$ vanishes. The analogous condition for $I_B$ is given by
\begin{align}
    (U \mp i\epsilon)(V \pm i\epsilon) = Z^2
\end{align}
which leads to the pole in $U$ (again for $+i\epsilon$) given by
$U_0 = \frac{Z^2}{V} - i\epsilon \frac{Z^2}{V^2} + i\epsilon$. This pole is inside the contour of integration of $U$ for $V^2 < Z^2$. As we there is also an integral over $V$, there will be some part of the integral that does not vanish. 
}

With this, we can write the two kernels as
\begin{align}
    z^{p}\mathbf{K}_{N} &= \frac{(-1)^{\frac{d-1}{2}}\Gamma(1 + \frac{d}{2} - p)2^{-p}}{2\pi^{\frac{d}{2}}\Gamma(1 - p)}\lim_{z'\rightarrow 0}(\sigma z')^{-p}\label{pk} \\
    z^{p}\mathbf{K}_{n N} &= \frac{(-1)^{\frac{d-1}{2}}\Gamma(1  - \frac{d}{2} + p)2^{p - d}}{2\pi^{\frac{d}{2}}\Gamma(1 - d + p)}\lim_{z' \rightarrow 0}(\sigma z')^{p - d}\label{dmpk}
\end{align}
Therefore, we arrive at the AdS-covariant form of the bulk reconstruction kernels for the modes of the massless $p$- form field. Formally, these kernels vanish due to the coefficient being $0$ for integer values of $p$. That said, this term must survive since these terms possess the correct exponents of $z$. One can argue that since the coefficient of the kernel is not unique, there are various possible arguments to recover a non-vanishing coefficient for it. As we shall see, Green's function approach also reproduces the same results. Moreover, we wish to mention that a similar problem was encountered in \cite{Kabat:2012hp}, where the denominator of the smearing function turned out to be divergent for a massive scalar with $\Delta\leq d-1$. Note that, we also have the same issue for free $p$-forms with the conformal dimension $\Delta=d-p$. Therefore, to avoid this divergence, one requires to follow the prescription given in the appendix (A) of \cite{Kabat:2012hp} and obtain a lightlike  smearing function which is supposed to be a delta function for $p>1$. We hope to investigate this in more detail in the future.

\subsection{Kernels under Hodge dual transformation} 
In $d+1$ dimensions, a $p$-form theory is Hodge dual to a $(d-p-1)$-form theory. As an example,
 a massless vector in three dimensions  can be dualized to a scalar by the relation
 $$F_{\alpha\beta}\sim\epsilon_{\alpha\beta}^{~\gamma}\partial_{\gamma}\phi.$$
 This duality equivalence holds true at the classical level. However, it is not obvious to hold it at the quantum level. It is observed that the logarithmic divergent term of $p$-forms which are Hodge dual to each other, do not have the same logarithmic divergence in the partition function \cite{Raj:2016zjp, David:2020mls, David:2021wrw}.

 Therefore, it is worth checking the properties of the kernel of $p$-forms under Hodge-duality. Under Hodge-dual transformation $p\rightarrow{ (d-p-1)}$,  and $\Delta\rightarrow (d + 1-\Delta)$. 

 Under the Hodge duality, we find that the kernels take the following form
 \begin{align}
     \mathbf{K}_{N} &\rightarrow \mathbf{\tilde{K}}_{N} = \frac{z^{2p + 2 -d }\Gamma(d/2)}{2\pi^{d/2 + 1}X^{d}}\;_{2}F_{1}\left(\frac{d}{2}, 1; 2-\frac{d}{2} +p; - \frac{z^2}{X^2}\right)\\
     \mathbf{K}_{nN} &\rightarrow \mathbf{\tilde{K}}_{nN} = \frac{\Gamma(d/2)}{2\pi^{d/2 + 1}X^{d}}\;_{2}F_{1}\left(\frac{d}{2}, 1;  - \frac{d}{2} + \Delta; - \frac{z^2}{X^2}\right)
 \end{align}
\begin{figure}[H]
  \centering
  \begin{minipage}[b]{0.4\textwidth}
    \includegraphics[width=\textwidth]{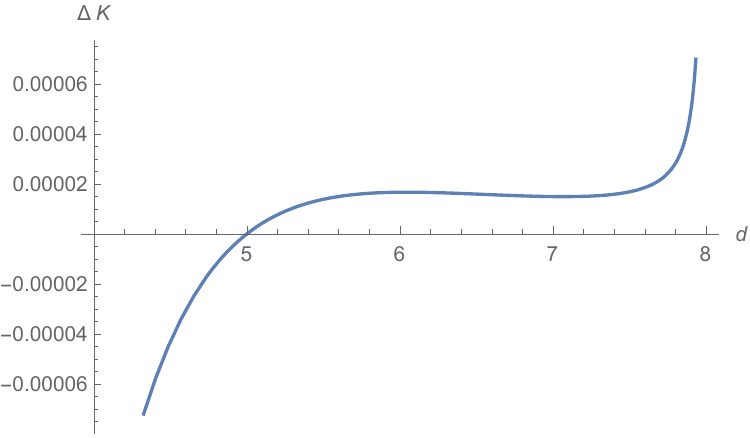}
    \caption{2-form.}
  \end{minipage}
  \hfill
  \begin{minipage}[b]{0.4\textwidth}
    \includegraphics[width=\textwidth]{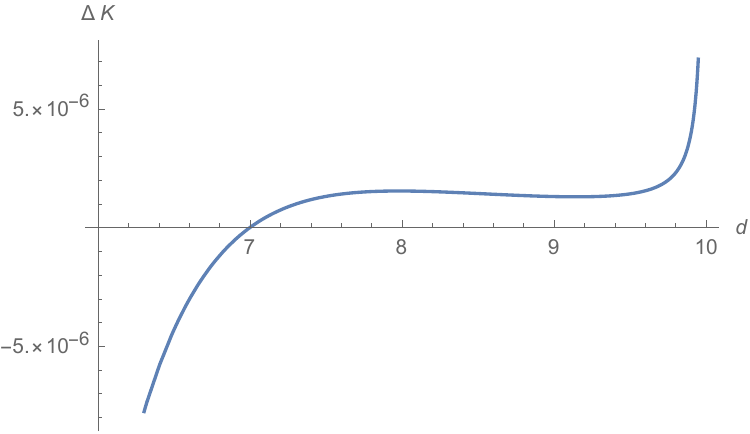}
    \caption{3-form.}\label{plot1}
  \end{minipage}
  \caption{We plot $\Delta K$, the difference between kernel of $2$-form and $3$-form and their Hodge dual as function of dimension. We set $z=1$ and $X=2$.}
\end{figure}
 Comparing these expressions to the kernels derived in \eqref{EvenAdSkernelp1} and \eqref{EvenAdSNNkernelp1}, we find that the kernels are identical for $d + 1 = 2\Delta$, which is the self-duality condition $2p = d - 1$. However, in general, the kernels differ under the Hodge-dual transformation. From plot \eqref{plot1}, we observe that the difference between the kernels of $2$-form and $3$-form and their Hodge dual vanishes only at the self-dual point. Since the duality relation is true at the classical level and therefore it is not obvious that a similar statement will also work for the smearing functions. But it will be interesting to reproduce the mismatch in the boundary partition from the difference of the smearing functions under Hodge dual transformation.

 \subsection{Spacelike kernel}
\begin{figure}[H]
  \centering
  \includegraphics[width=0.5\textwidth]{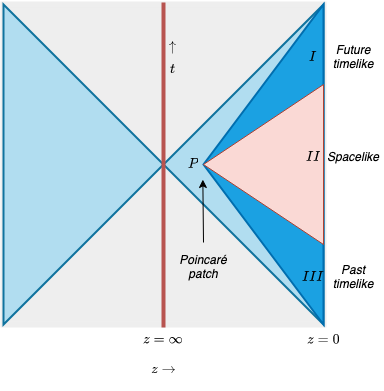}
  \caption{Penrose diagram of empty AdS$_{d + 1}$ spacetime. The bulk point at which the fields are reconstructed is denoted by $P$, while the spacelike, future timelike, and past timelike regions of $P$ are denoted by $I$, $II$, and $III$, respectively. The AdS boundary is indicated by $z = 0$.}\label{Figfig}
\end{figure}

 In this section, we restrict the kernels to the spacelike region of the bulk point at which the field is reconstructed. This is expected since the HKLL bulk reconstruction is a causal reconstruction procedure. We begin by focusing on the kernel for the normalizable mode \eqref{pk}. The bulk field constructed from this kernel at a bulk point $P$ is given as (the $\lim_{z'\rightarrow 0}$ is implicit)
 \begin{align}
     \Psi_{J}(z, x; x') = \int \mathrm{d}t'\mathrm{d}\vec{x}' \alpha_{d}(\sigma z')^{-p}\phi_{J}^{(0)}(x')
 \end{align}
 where we denote $\alpha_{d} =  \frac{(-1)^{\frac{d}{2}}\Gamma(1 + \frac{d}{2} - p)2^{\Delta - d}}{2\pi^{\frac{d}{2}}\Gamma(1 - p)}$. Here we define $\Psi_{J} = z^{p}A_{J}$. The integral is over the full AdS boundary, and we seek to restrict it to the spacelike separated region. For this purpose, we introduce the following redefinition of the boundary fields $\phi_{J, 0}(x')$
 \begin{align}
     \phi_{J}^{(0)}(x') = \begin{cases}
         &-e^{-i \pi p}\tilde{\phi}_{J}^{(0)}(x')\;\;\text{future timelike region of P}(\mathbf{I}) \\
         &\tilde{\phi}_{J}^{(0)}(x')\;\;\;\;\;\;\;\;\;\;\;\;\;\text{spacelike region of P} (\mathbf{II}) \\
         &-e^{i \pi p}\tilde{\phi}_{J}^{(0)}(x')\;\;\;\;\text{past timelike region of P}(\mathbf{III})
     \end{cases}
 \end{align}
 With this redefinition, we write the bulk field as
 \begin{align}
     \Psi_{J}(z, x; x') = \int \mathrm{d}t'\mathrm{d}\vec{x}' \alpha_{d}|\sigma z'|^{-p}\begin{cases}
         &-e^{-i \pi p}\tilde{\phi}_{J}^{(0)}(x')\;\;\text{future timelike region of P}(\mathbf{I}) \\
         &\tilde{\phi}_{J}^{(0)}(x')\;\;\;\;\;\;\;\;\;\;\;\;\;\text{spacelike region of P} (\mathbf{II}) \\
         &-e^{i \pi p}\tilde{\phi}_{J}^{(0)}(x')\;\;\;\;\text{past timelike region of P}(\mathbf{III})
     \end{cases}
 \end{align}
 For a diagrammatic representation of the different regions within the Poincar\'e patch, see Fig\,.\ref{Figfig}.
 
 To this field $\Psi_{J}$, one can add the following term
 \begin{align}
     F = \left(\frac{1}{2 z}(z^2 + |\vec{x} - \vec{x'}|^2 - (t' - t - i\epsilon)^2)\right)^{-p}
 \end{align}
 This term vanishes when integrated against the boundary field. In the future and past timelike regions, this term picks up the following phases
 \begin{align}
     F = \begin{cases}
     &e^{-i \pi p}|\sigma z'|^{-p}\;\;\;\;\;\text{future timelike region of P}(\mathbf{I}) \\
     &|\sigma z'|^{-p}\;\;\;\;\;\;\;\;\;\;\;\;\;\text{spacelike region of P} (\mathbf{II}) \\
     &e^{i \pi p}|\sigma z'|^{-p}\;\;\;\;\;\;\text{past timelike region of P}(\mathbf{III})
     \end{cases}
 \end{align}
 This term, when added to the kernel, cancels out in the past and future timelike regions and gives a factor of $2$ in the spacelike region. The redefined kernel is as follows
 \begin{align}
     \mathbf{K}_{N} \rightarrow \tilde{\mathbf{K}}_{N} = \mathbf{K}_{N} + \alpha_{d}F
 \end{align}
 With this redefinition, the bulk normalizable field can be written as
 \begin{align}
     \Psi_{J}(z, x; x') = 2\int \mathrm{d}t'\mathrm{d}\vec{x}' \alpha_{d}(\sigma z')^{-p}\theta(\text{spacelike})\phi_{J}^{(0)}(x')
 \end{align}
 Thus the kernel can be read off as
 \begin{align}
     \tilde{\mathbf{K}}_{N} = \frac{(-1)^{\frac{d}{2}}\Gamma(1 + \frac{d}{2} - p)2^{-p}}{\pi^{\frac{d}{2}}\Gamma(1 - p)}\lim_{z' \rightarrow 0}(\sigma z')^{-p}\theta(\text{spacelike})\label{KNpFormMDSUM}
 \end{align}
 where we have used the value of $\alpha_{d}$ and reintroduced the limit. In a similar way, the non-normalizable mode can be studied. The bulk non-normalizable mode is written as (with the non-normalizable kernel from \eqref{dmpk})
 \begin{align}
     \Psi_{J}(z, x; x') = \int \mathrm{d}t' \mathrm{d}^{d}\vec{x}' \beta_{d}(\sigma z')^{p-d}j_{J}^{(0)}(x')
 \end{align}
 where $\beta_{d} =  \frac{(-1)^{\frac{d}{2}}\Gamma(1  - \frac{d}{2} + p)2^{p - d}}{2\pi^{\frac{d}{2}}\Gamma(1 - d + p)}$. The boundary field $j_{J, 0}$ can be redefined in a similar manner to the  
 \begin{align}
     j_{J}^{(0)} = \begin{cases}
         &e^{i\pi p}\tilde{j}_{J}^{(0)}(x')\;\;\;\;\;\;\;\;\text{future timelike region of P}(\mathbf{I})\\
         &\tilde{j}_{J}^{(0)}(x')\;\;\;\;\;\;\;\;\;\;\;\;\;\;\text{spacelike region of P} (\mathbf{II})\\
         &e^{-i\pi p}\tilde{j}_{J}^{(0)}(x')\;\;\;\;\;\;\text{past timelike region of P}(\mathbf{III})
     \end{cases}
 \end{align}
 With this redefinition, the bulk non-normalizable mode is written as 
 \begin{align}
     \Psi_{J}(z, x; x') = \int \mathrm{d}t' \mathrm{d}^{d}\vec{x}' \beta_{d}(\sigma z')^{p-d}\begin{cases}
         &e^{i\pi p}\tilde{j}_{J}^{(0)}(x')\;\;\;\;\;\;\;\;\text{future timelike region of P}(\mathbf{I})\\
         &\tilde{j}_{J}^{(0)}(x')\;\;\;\;\;\;\;\;\;\;\;\;\;\;\text{spacelike region of P} (\mathbf{II})\\
         &e^{-i\pi p}\tilde{j}_{J}^{(0)}(x')\;\;\;\;\;\;\text{past timelike region of P}(\mathbf{III})
     \end{cases}
 \end{align}
 To this kernel, we add the following function
 \begin{align}
     F' = \left(\frac{1}{2 z}(z^2 + |\vec{x} - \vec{x'}|^2 - (t' - t - i\epsilon)^2)\right)^{p-d}
 \end{align}
 This function takes the following form in the timelike and spacelike regions
 \begin{align}
     F' = \begin{cases}
     &-e^{i \pi p}|\sigma z'|^{p-d}\;\;\;\;\;\;\;\;\text{future timelike region of P}(\mathbf{I}) \\
     &|\sigma z'|^{p-d}\;\;\;\;\;\;\;\;\;\;\;\;\;\;\;\;\;\text{spacelike region of P} (\mathbf{II}) \\
     &-e^{-i \pi p}|\sigma z'|^{p-d}\;\;\;\;\;\;\text{past timelike region of P}(\mathbf{III})
     \end{cases}
 \end{align}
 This function can be added to the kernel $\mathbf{K}_{n N}$ to redefine the kernel as follows
 \begin{align}
     \mathbf{K}_{n N} \rightarrow \tilde{\mathbf{K}}_{n N} = \mathbf{K}_{n N} + \beta_{d}F'
 \end{align}
 This kernel, therefore, takes the following spacelike form
 \begin{align}
     \tilde{\mathbf{K}}_{n N} =  \frac{(-1)^{\frac{d}{2}}\Gamma(1  - \frac{d}{2} + p)2^{p - d}}{\pi^{\frac{d}{2}}\Gamma(1 - d + p)}\lim_{z' \rightarrow 0}(\sigma z')^{p-d}\theta(\text{spacelike})\label{KnNpFormMDSUM}
 \end{align}
where we have used the value of $\beta_{d}$ and reintroduced the limit. {\color{black}Therefore we restrict the mode-sum kernels to the spacelike region of the bulk point.  Note that, the gauge potential is defined as $\Psi=z^p A_J$, and therefore, the smearing functions \eqref{KNpFormMDSUM} and \eqref{KnNpFormMDSUM}  agree with the smearing functions obtained using mode sum approach in \eqref{pk} and \eqref{dmpk}.}
\subsection{Lighlike smearing function}
 Apparently, the smearing functions in \eqref{pk} and \eqref{dmpk} are divergent for specific $p$-forms. To circumvent this issue, we follow \cite{Kabat:2012hp}, and construct light-like smearing functions. Since, we are considering massless $p$-form fields, it is best to work with lightlike smearing functions.
 \begin{align}\label{lls}
     z^p A_p(x,x;x')=\Psi_J(z,x;x')&=\mathcal{N}\int_{t'^2+|\vec{x'}|^2=z^2} dt' d\vec{x'}\, \phi^{(0)}_J(t+t', \vec{x}+i\vec{x'})\nonumber\\&=\mathcal{N}\int dt'd\vec{x'}\delta(\sigma z')\phi^{(0)}_J(t+t', \vec{x}+i\vec{x'}), \quad\quad\mathcal{N}=\frac{1}{\rm{vol}(S^{d-1})},
 \end{align}
 where $(\sigma z')$ is the covariant distance in $AdS$.
 After splitting the boundary coordinates  $x' = (t'; \vec{x'})$ into a time coordinate $t'$ and spatial coordinates $\vec{x'}$, we evaluate the boundary current at complex values of the spatial coordinates. The boundary integral is performed over
a sphere of radius $z$ on the complexified boundary. The centre  of the 
sphere located at $(t, \vec{x})$. 

Now we need to check the gauge field in \eqref{lls} satisfies the boundary condition and the equation of motion. Checking the boundary condition is reasonably simple. It is clear that in the limit $z\rightarrow 0$, the region of the integration (defined in the limit of \eqref{lls}) shrinks to a point and we can take out $\phi^{(0)}_J$ out of the integral. The rest of the integral evaluates to the volume of boundary sphere which cancels with $\mathcal{N}$ and the boundary condition is satisfied.

To check the equation of motion, we act with the wave operator and finally integrate against a test function $f(\sigma z')$. Note that, when $z'$ is small, the smearing function in \eqref{lls} can be written as simply, $\frac{\delta(\sigma)}{z'}$, and we wish to check that this is annihilated by the wave operator in the limit $z'\rightarrow 0$. Let us proceed with the wave operator of the $p$-forms 
\begin{align}\label{lls1}
    (\sigma^2 - 1)\frac{d^2}{d\sigma^2}\Psi_{J}(\sigma) + (d  + 1)\sigma \frac{d}{d\sigma}\Psi_{J}(\sigma) - \Delta(\Delta - d)\Psi_{J}(\sigma) = 0
\end{align}
Substituting, $\Delta=d-p$, and integrating against a test function $f(\sigma z')$ in \eqref{lls1}, we find
\begin{align}\label{lls2}
    &\int d(\sigma z') f(\sigma z') \left[ (\sigma^2 - 1) \frac{d^2}{d\sigma^2} + (d+1)\sigma \frac{d}{d\sigma} + p(d-p) \right] \frac{1}{z'} \delta(\sigma)\nonumber\\
    &= \int d(\sigma z') \frac{1}{z'} \delta(\sigma) \left[ \frac{d^2}{d\sigma^2}(\sigma^2 - 1) - (d+1) \frac{d}{d\sigma} \sigma +p (d-p) \right] f(\sigma z')\nonumber\\
    &=-z'^2f''(0)
\end{align}
In the second line, we have integrated by parts and then perform the integral with the delta function against the test function. It is clear that, in the limit $z'\rightarrow 0$, the equation of motion is satisfied. Therefore, the smearing function in \eqref{lls} satisfies both the boundary condition as well as the equation of motion.
\section{Graviton}\label{g1}
 We now focus our attention on deriving HKLL kernels corresponding to the two modes of the graviton. The gauge redundancy is realized through the diffeomorphism in gravity. The Einstein-Hilbert action is given by
 \begin{align}
     S&=\frac{1}{4G_N}\int d^{d+1}x(R-2\Lambda), \quad\quad \Lambda=-\frac{d(d-1)}{2}.
 \end{align}
 We work with the linearized order in metric perturbations around the AdS
background
 $$g_{MN}=\bar{g}_{MN}+h_{MN}.$$
 The  bulk equation of linearized  graviton in AdS space is obtained as \cite{Sarkar:2014dma}
\begin{eqnarray} \label{laggrav}
\nabla_L\nabla^Lh_{MN}-2\nabla_L\nabla_Mh^L_{~N}+\nabla_M\nabla_N h^L_{~L}-2dh_{MN}=0.
\end{eqnarray}
We will obtain the bulk equation of the linearized graviton on the Poincar\'e patch of AdS.
 \subsection{Bulk wave equation}
 We wish to evaluate the bulk equation corresponding to the transverse traceless degrees of freedom. Therefore, we impose the following gauge restrictions 
 \begin{align}
     h^{\alpha}_{\alpha}=0,\quad\quad \partial_{\mu}h^{\mu}_{\nu}=0.
 \end{align}
 These gauge conditions will imply the conservation of currents and the tracelessness of the stress tensor 
 \begin{align}
     \partial_{\mu}T^{\mu\nu}=0,\quad\quad T^{\mu}_{\mu}=0.
 \end{align}
 We also work with the holographic gauge to remove residual gauge invariances \cite{Sarkar:2014jia}
 \begin{align}
     h_{zz}=0,\quad\quad h_{z\mu}=0.
 \end{align}
 We  now impose the gauge conditions and obtain the bulk wave equation
 of the graviton in the Poincar\'e patch of AdS 
\begin{align}
    \partial_\alpha \partial^\alpha h_{\mu \nu} + \partial^{2}_{z}h_{\mu \nu} + \frac{5-d}{z}\partial_{z}h_{\mu \nu} - \frac{2(d-2)}{z^2}h_{\mu \nu} = 0
\end{align}
At this stage it is convenient to substitute $\Phi_{\mu \nu} = z^2 h_{\mu \nu}$ and the rescaled bulk equation becomes
\begin{align}
    \partial_{\alpha}\partial^\alpha \Phi_{\mu \nu} + z^{d-1}\partial_{z}
\left(z^{1-d}\partial_{z}\Phi_{\mu \nu} \right) = 0\label{phimunueqn}
\end{align}
We will later show that this rescaled wave equation admits AdS covariant form.
The solution to the wave equation in this background is obtained as
\begin{align}
     \Phi_{\mu \nu}(z, x) = \int_{|q| \geq 0}a_{\mu \nu}(q)z^{d/2}J_{d/2}(q z)e^{i q . x}\frac{\mathrm{d}^{d}q}{(2\pi)^{d}} +\int_{|q| \geq 0}b_{\mu \nu}(q)z^{d/2}Y_{d/2}(q z)e^{i q . x}\frac{\mathrm{d}^{d}q}{(2\pi)^{d}}\label{gravsoln}
\end{align}
Therefore, the mode expansion of the bulk field $h_{\mu\nu}(z,x)$ can be expressed as
\begin{align}
     h_{\mu \nu}(z, x)= \int_{|q| \geq 0}a_{\mu \nu}(q)z^{\frac{d}{2}-2}J_{\frac{d}{2}}(q z)e^{i q . x}\frac{\mathrm{d}^{d}q}{(2\pi)^{d}} +\int_{|q| \geq 0}b_{\mu \nu}(q)z^{\frac{d}{2}-2}Y_{\frac{d}{2}}(q z)e^{i q . x}\frac{\mathrm{d}^{d}q}{(2\pi)^{d}}\label{gravsoln}
\end{align}
Here $x\equiv (t,\Vec{x})$ and $q \equiv (\omega, \Vec{k})$ with $q=\sqrt{(\omega^2-|k|^2}.$ The near boundary behavior of the solution can be extracted from the expansion of the Bessel functions at $z=0$. For now, we focus on the case where $\frac{d}{2}-2\neq \text{Integer}$. This corresponds to the case where $d$ is odd, i.e., Even AdS.
\subsection{Even AdS}
In even dimensional AdS\footnote{The odd AdS case is discussed in Appendix \ref{appC}}, there are two independent solutions to the bulk wave equation, which are $J_{\nu}(|q|z)$ and $J_{-\nu}(|q|z)$ which is due to the relation given in \eqref{brel}. Therefore the solution to the wave equation in this background can be expressed.
\begin{align}
     h_{\mu\nu}(z, x) = \int_{|q| \geq 0}\frac{\mathrm{d}^{d}q}{(2\pi)^{d}}z^{\frac{d}{2}-2}\Big[a_{\mu \nu}(q)J_{\frac{d}{2}}(q z)+b_{\mu\nu}(q)J_{-\frac{d}{2}}(qz)\Big]e^{i q . x} 
\end{align}
From the above solution, we read off the mode solutions for the case of even AdS.
\begin{align}
    h_{\mu \nu}(z,x)\Big\vert_{N} &= \int_{|q| \geq 0}a_{\mu \nu}(q)z^{\frac{d}{2}-2}J_{d/2}(q z)e^{i q . x}\frac{\mathrm{d}^{d}q}{(2\pi)^{d}} = \sum_{k = 0}^{\infty}z^{d + 2 k-2}\phi^{(k)}_{\mu \nu}(x)\label{norm1} \\
    h_{\mu \nu}(z,x)\Big\vert_{n-N} &= \int_{|q| \geq 0}b_{\mu \nu}(q)z^{\frac{d}{2}-2}J_{-d/2}(q z)e^{i q . x}\frac{\mathrm{d}^{d}q}{(2\pi)^{d}} = \sum_{k=0}^{\infty}z^{2 k-2}j^{(k)}_{\mu \nu}(x) \label{nonnorm1}
\end{align}
The coefficients at each order can be expressed
\begin{align}\label{intgrav}
    \phi^{(k)}_{\mu \nu}(x) &= \frac{(-1)^{k}2^{-2k - \frac{d}{2}}}{\Gamma(k+1)\Gamma(k + \frac{d}{2}+1)}\int_{q \geq 0}a_{\mu \nu}(q)q^{2 k + \frac{d}{2}}e^{i q. x}\frac{\mathrm{d}^{d}q}{(2\pi)^{d}} \\
    j^{(k)}_{\mu \nu}(x) &= \frac{(-1)^{k}2^{-2k + \frac{d}{2}}}{\Gamma(k+1)\Gamma(k - \frac{d}{2}+1)}\int_{q \geq 0}b_{\mu \nu}(q)q^{2 k - \frac{d}{2}}e^{i q. x}\frac{\mathrm{d}^{d}q}{(2\pi)^{d}}
\end{align}
\subsection*{Kernels as mode-sum integrals}
The integrals \eqref{intgrav} can be inverted to obtain $a_{\mu\nu}(|q|)$ and $b_{\mu\nu}(|q|)$. To do that, one can pick two independent data from the expansion coefficients of the two modes. We choose coefficients corresponding to $k=0$
and interpret as holographic data.
\begin{align}
    \phi^{(0)}_{\mu \nu}(x) &= \frac{2^{-\frac{d}{2}}}{\Gamma(\frac{d}{2}+1)}\int_{|q| \geq 0} a_{\mu \nu}(q)q^{\frac{d}{2}}e^{i q. x}\frac{\mathrm{d}^{d}q}{(2\pi)^{d}} \label{phimunu}\\
    j^{(0)}_{\mu \nu}(x) &= \frac{2^{\frac{d}{2}}}{\Gamma(1-\frac{d}{2})}\int_{|q| \geq 0} b_{\mu \nu}(q)q^{-\frac{d}{2}}e^{i q. x}\frac{\mathrm{d}^{d}q}{(2\pi)^{d}} \label{jmunu}
\end{align}

Inverting this relation, we obtain
\begin{align}
    \int \phi^{(0)}_{\mu \nu}(x)e^{-i k. x}\mathrm{d}^{d}x =\frac{2^{-\frac{d}{2}}}{\Gamma(\frac{d}{2}+1)}a_{\mu \nu}(k)k^{\frac{d}{2}}
\end{align}
Inserting this expression for $a_{\mu \nu}(q)$ into the expression \eqref{norm1}, we find the following expression
\begin{align}
    h_{\mu \nu}(z, x)|_{N} = z^{\frac{d}{2}-2}2^{\frac{d}{2}}\Gamma(\frac{d}{2}+1)\int_{|q|\geq 0} \frac{J_{d/2}(q z)}{q^{d/2}}e^{i q. (x - x')}\phi^{(0)}_{\mu \nu}(x')\frac{\mathrm{d}^{d}q \mathrm{d}^{d}x'}{(2\pi)^{d}}
\end{align}
The kernel is defined as 
\begin{align}
    h_{\mu \nu}(z, x)|_{N} = \int \mathbf{K}_{N}(z, x; x') \phi^{(0)}_{\mu \nu}(x')\mathrm{d}^{d}x'
\end{align}
From this, we can read off the kernel for the normalizable mode
\begin{align}\label{gravN}
    \mathbf{K}_{N} = z^{\frac{d}{2}-2}2^{\frac{d}{2}}\Gamma(\frac{d}{2}+1)\int_{|q| \geq 0}q^{-\frac{d}{2}}J_{\frac{d}{2}}(q z)e^{i q. (x - x')}\frac{\mathrm{d}^{d}q}{(2\pi)^{d}}
\end{align}
Let us now focus on the non-normalizable mode.
\begin{align}
    j^{(0)}_{\mu \nu}(x) =  \frac{2^{\frac{d}{2}}}{\Gamma(1-\frac{d}{2})}\int_{|q| \geq 0}q^{-\frac{d}{2}}b_{\mu \nu}(q)e^{i q.x}\frac{\mathrm{d}^{d}q}{(2\pi)^{d}}
\end{align}
 Inverting this relation, we get
\begin{align}
    \int j_{\mu \nu}(x)e^{-i k. x}\mathrm{d}^{d}x =  \frac{2^{\frac{d}{2}}}{\Gamma(1-\frac{d}{2})}k^{-\frac{d}{2}}b_{\mu \nu}(k) 
\end{align}
Plugging this expression for $b_{\mu\nu}(q)$ into \eqref{nonnorm1}, we find the following expression
\begin{align}
    h_{\mu \nu}(z, x)|_{n-N} = z^{\frac{d}{2}-2}2^{-\frac{d}{2}}\Gamma(1-\frac{d}{2})\int_{|q| \geq 0} J_{\frac{d}{2}}(k z)k^{d/2}e^{i k. (x - x')}j_{\mu \nu}(x)\frac{\mathrm{d}^{d}k\mathrm{d}^{d}x}{(2\pi)^d}
\end{align}
From this, we can read off the kernel expression as 
\begin{align}\label{gravnN}
    \mathbf{K}_{nN} = z^{\frac{d}{2}-2}2^{-\frac{d}{2}}\Gamma(1-\frac{d}{2})\int_{|q| \geq 0}J_{-\frac{d}{2}}(k z)q^{d/2}e^{i q.(x-x')}\frac{\mathrm{d}^{d}q}{(2\pi)^{d}}
\end{align}
\subsection*{Explicit evaluation of the Poincar\'e kernel integrals}
In this section, we evaluate the kernels \eqref{gravN} and \eqref{gravnN} in the Poincar\'e patch of the even-dimensional AdS space.  We use the integral given in \eqref{identity_integral-0}
\begin{align}
    \int_{|q| \geq 0}\frac{\mathrm{d}^{d}q}{(2\pi)^{d}}|q|^{\mu}\zeta_{\nu}(q z)e^{i q. (x - x')} = \frac{1}{\pi(2\pi)^{d/2}X^{d/2 - 1}}\int_{x = 0}^{\infty}x^{\mu + d/2}\zeta_{\nu}(x z)K_{\frac{d}{2}-1}(x X)\mathrm{d}x 
\end{align}
where $\zeta$ is some function of $q z$, while $\mu$ and $\nu$ are some complex numbers. We begin with considering \eqref{gravN}, where $\mu = -\frac{d}{2}$ and $\zeta_{\nu}(qz) = J_{\frac{d}{2}}(qz)$. This gives us the following integral
\begin{align}
    \mathbf{K}_{N} =2^{\frac{d}{2}}\Gamma(\frac{d}{2}+1)\frac{z^{\frac{d}{2}-2}}{X^{d/2 - 1}}\int_{x = 0}^{\infty}J_{\frac{d}{2}}(z x)K_{\frac{d}{2}-1}(x X)\mathrm{d}x\label{Normkernel1}
\end{align}
To evaluate this integral, we need to use the following identity
\begin{align}
    \int_{0}^{\infty}x^{-\lambda}K_{\mu}(a x)J_{\nu}(b x) &= \frac{b^{\nu}\Gamma(\frac{\nu -\lambda + \mu + 1}{2})\Gamma(\frac{\nu -\lambda - \mu + 1}{2})}{2^{\lambda + 1}\Gamma(\nu + 1)a^{-\lambda + \nu + 1}}\notag\\
    &\times\;_{2}F_{1}\left(\frac{\nu -\lambda + \mu + 1}{2}, \frac{\nu -\lambda - \mu + 1}{2}; \nu + 1; -\frac{b^2}{a^2}\right)\label{identity_integral-1}
\end{align}
which holds for $\text{Re}(\nu + 1 -\lambda) > \vert\text{Re}(\mu)\vert$ and $\text{Re}(a \pm i b)$. In the integral \eqref{Normkernel1}, we have $\mu = \frac{d}{2}-1$, $\nu = \frac{d}{2}$ and $\lambda = 0$. Thus we have, with $a = X$ and $b = z$
\begin{align}
    \mathbf{K}_{N} = \frac{\Gamma(\frac{d}{2})}{2 \pi^{d/2 + 1}}\frac{z^{d-2}}{X^{d}}\;_{2}F_{1}\left(\frac{d}{2}, 1; \frac{d}{2} + 1; -\frac{z^2}{X^2}\right)\label{normKerGraviton}
\end{align}
We turn to evaluate the non-normalizable kernel integral \eqref{gravnN}. From the same integral identity \eqref{identity_integral-0} we identify $\zeta_\nu (qz)= J_{-\frac{d}{2}}(qz)$, $\nu = -\frac{d}{2}$, $\lambda = -d$ and $\mu = \frac{d}{2}$). Therefore, the integral becomes
\begin{align}
    \mathbf{K}_{nN} = \frac{2^{-\frac{d}{2}}\Gamma(1-\frac{d}{2})}{\pi  (2\pi)^{d/2}}\frac{z^{\frac{d}{2}-2}}{X^{d/2 - 1}}\int_{x = 0}^{\infty}x^{d}J_{-\frac{d}{2}}(z x)K_{\frac{d}{2}-1}(z X)\mathrm{d}x\label{nonNormkernel1}
\end{align}
To evaluate this integral, we use the identity \eqref{identity_integral-1} again. The conditions are automatically satisfied. This gives the following expression (with $\lambda = -d$, $\mu = \frac{d}{2}$, $\nu = \frac{d}{2}-1$, $a = z$ and $b = X$)
\begin{align}
    \mathbf{K}_{nN} = \frac{\Gamma(d/2)z^{-2}}{2 \pi^{d/2 + 1}X^{d}}\;_{2}F_{1}\left(1, \frac{d}{2}; 1 - \frac{d}{2}; -\frac{z^2}{X^2}\right)\label{nonnormKerGraviton}
\end{align}

\subsection{Covariant part of the kernels}
In this section, we use transformation properties of the hypergeometric $\;_{2}F_{1}$ function in order to extract the AdS-covariant piece from the kernels for the two modes. 

The particular transformation that we utilize is the following
\begin{align}
    \;_{2}F_{1}(a, b; c; z) = &\frac{\Gamma(c)\Gamma(b - a)}{\Gamma(b)\Gamma(c-a)}(-z)^{-a}\;_{2}F_{1}\left(a, 1 - c + a; 1 - b + a; \frac{1}{z}\right) \notag\\ & + \frac{\Gamma(c)\Gamma(a - b)}{\Gamma(a)\Gamma(c-b)}(-z)^{-b}\;_{2}F_{1}\left(b, 1 - c + b; 1 - a + b; \frac{1}{z}\right)
\end{align}
This lets us rewrite the kernel expression\eqref{normKerGraviton} as
\begin{align}
    z^{2}\mathbf{K}_{N} = \frac{\Gamma(\frac{d}{2} + 1)\Gamma(\frac{d}{2} - 1)}{\Gamma(\frac{d}{2})2 \pi^{\frac{d}{2} + 1}}\frac{z^{d-2}}{X^{d-2}}\;_{2}F_{1}\left( 1, 1 - \frac{d}{2}; 2 - \frac{d}{2}; - \frac{X^2}{z^2}\right) + \frac{\Gamma(\frac{d}{2})\Gamma(\frac{d}{2}+1)\Gamma(1 - \frac{d}{2})}{2\pi^{\frac{d}{2} + 1}}\label{covGrav1}
\end{align}
And similarly \eqref{nonnormKerGraviton} as
\begin{align}
    z^{2}\mathbf{K}_{n N} = \frac{\Gamma(1 - \frac{d}{2})\Gamma(\frac{d}{2} - 1)z^{-2}}{2 \pi^{\frac{d}{2} + 1} X^{d - 2}}\;_{2}F_{1}\left(1; 1 + \frac{d}{2}; 2 - \frac{d}{2}; - \frac{X^2}{z^2} \right)+ \frac{(-1)^{\frac{d}{2} - 1}2^{-d}\Gamma(1 - \frac{d}{2})}{2 \pi^{d/2}\Gamma(1 - d)}\lim_{z'\rightarrow 0}(\sigma z')^{-d}\label{covGrav2}
\end{align}
As can be seen from these expressions, the second terms in \eqref{covGrav1} and \eqref{covGrav2} are the AdS-covariant pieces with the correct exponents. The first terms in both these expressions do not contain any piece with the same exponent in $z$ as the second term. Therefore these terms are distinct, and there is no mixing between them.

We note that the first term in \eqref{covGrav1} only has terms proportional to $z^{d - 2 - 2n}$. Since such terms do not arise in \eqref{norm1} (once the $z^2$ scaling is taken care of), this term should vanish when integrated against the boundary field. Similarly, the first term in \eqref{covGrav2} only has terms proportional to $z^{-2 -2n}$. Terms like these do not arise in \eqref{nonnorm1}, and therefore the first term in \eqref{covGrav2} should also vanish when integrated against the boundary field.

By the prescription discussed in Appendix D of \cite{Bhattacharjee:2022ehq}, both of the first terms in \eqref{covGrav1} and \eqref{covGrav2} can be dropped since $X = 0$ is a pole/branch-point. Hence, with that, we are left with the following expressions of the kernels
\begin{align}
    z^{2}\mathbf{K}_{N} &= \frac{\Gamma(\frac{d}{2})\Gamma(\frac{d}{2}+1)\Gamma(1 - \frac{d}{2})}{2\pi^{\frac{d}{2} + 1}} \\
    z^{2}\mathbf{K}_{n N} &= \frac{(-1)^{\frac{d}{2}-1}2^{-d}\Gamma(1 - \frac{d}{2})}{2 \pi^{d/2}\Gamma(1 - d)}\lim_{z'\rightarrow 0}(\sigma z')^{-d}
\end{align}
Therefore, we arrive at the expressions for the bulk reconstruction kernels corresponding to the two modes of the graviton.

\subsection{Spacelike kernel}
In this section, we restrict the mode-sum kernels to the spacelike region of the bulk point. We begin by considering the non-normalizable mode. The kernel reconstructing the bulk non-normalizable mode is written as (the $\lim_{z' \rightarrow 0}$ is implicit)
\begin{align}
    \Phi_{\mu \nu}(z, x; x') = \int \mathrm{d}t' \mathrm{d}^{d - 1}\vec{x'} a_{d}(\sigma z')^{-d}j^{(0)}_{\mu \nu}(x')
\end{align}
where $a_{d} = \frac{(-1)^{\frac{d}{2} - 1}2^{-d}\Gamma(1 - \frac{d}{2})}{2 \pi^{d/2}\Gamma(1 - d)}$. To restrict this kernel to the spacelike separated boundary region of the bulk point $P$ (where the bulk field is reconstructed), we redefine the boundary field as follows
\begin{align}
j^{(0)}_{\mu \nu}(x') = \begin{cases}
         &-e^{-i \pi d}\tilde{j}^{(0)}_{\mu \nu}(x')\;\;\text{future timelike region of P}(\mathbf{I}) \\
         &\tilde{j}^{(0)}_{\mu \nu}(x')\;\;\;\;\;\;\;\;\;\;\;\;\;\text{spacelike region of P} (\mathbf{II}) \\
         &-e^{i \pi d}\tilde{j}^{(0)}_{\mu \nu}(x')\;\;\;\;\text{past timelike region of P}(\mathbf{III})
     \end{cases}
\end{align}
However, since $d$ is odd, the overall coefficient is $+1$, and so the redefined field is identical to the original field $j^{(0)}_{\mu \nu}(x') = \tilde{j}^{(0)}_{\mu \nu}$. To the kernel $z^2\mathbf{K}_{n N} = \alpha_{d}|\sigma z'|^{-d}$ we add the following function
\begin{align}
F = \left( \frac{1}{2 z}(z^2 + |x - x'|^2 - (t' - t - i\epsilon)^2)\right)^{-d}
\end{align}
This vanishes when integrated against $\phi^{(0)}_{\mu \nu}(x')$.

In the timelike and spacelike region of $P$, this function takes the following form
\begin{align}
     F = \begin{cases}
     &-|\sigma z'|^{-d}\;\;\;\;\;\;\;\;\text{future timelike region of P}(\mathbf{I}) \\
     &|\sigma z'|^{p-d}\;\;\;\;\;\;\;\;\;\text{spacelike region of P} (\mathbf{II}) \\
     &-\sigma z'|^{-d}\;\;\;\;\;\;\;\;\;\text{past timelike region of P}(\mathbf{III})
     \end{cases}
\end{align}
It follows from this expression that the redefined kernel
\begin{align}
    \tilde{\mathbf{K}}_{n N} = \mathbf{K}_{n N} + a_{d}F
\end{align}
vanishes in the timelike region and gives a factor of $2$ in the spacelike region of $P$. Hence the effective non-normalizable kernel is as follows
\begin{align}
    \tilde{\mathbf{K}}_{n N} = \frac{(-1)^{\frac{d}{2} - 1}2^{-d}\Gamma(1 - \frac{d}{2})}{\pi^{d/2}\Gamma(1 - d)}\lim_{z' \rightarrow 0}(\sigma z')^{-d}\theta(\text{spacelike})\label{jejejeje2}
\end{align}
where we have used the value of $a_{d}$ and reintroduced the limit. Now we consider the normalizable mode. The normalizable bulk field is written as
\begin{align}
    \Phi_{\mu \nu}(z, x; x') = \int \mathrm{d}t'\mathrm{d}^{d - 1}\vec{x}'b_{d}\phi^{(0)}_{\mu \nu}(x')
\end{align}
where $b_{d} = \frac{\Gamma(\frac{d}{2})\Gamma(\frac{d}{2}+1)\Gamma(1 - \frac{d}{2})}{2\pi^{\frac{d}{2} + 1}}$. We redefine the boundary field as follows
\begin{align}
    \phi^{(0)}_{\mu \nu}(x') = \begin{cases}
         &-\tilde{\phi}^{(0)}_{\mu \nu}(x')\;\;\;\;\;\;\;\;\;\;\text{future timelike region of P}(\mathbf{I}) \\
         &\tilde{\phi}^{(0)}_{\mu \nu}(x')\;\;\;\;\;\;\;\;\;\;\;\;\;\text{spacelike region of P} (\mathbf{II}) \\
         &-\tilde{\phi}^{(0)}_{\mu \nu}(x')\;\;\;\;\;\;\;\;\;\;\text{past timelike region of P}(\mathbf{III})
     \end{cases}
\end{align}
Therefore the bulk normalizable mode is written as
\begin{align}
    \Phi_{\mu \nu}(z, x; x') = \int \mathrm{d}t' \mathrm{d}^{d - 1}\vec{x}' b_{d}\begin{cases}
         &-\tilde{\phi}^{(0)}_{\mu \nu}(x')\;\;\;\;\;\;\;\;\;\;\text{future timelike region of P}(\mathbf{I}) \\
         &\tilde{\phi}^{(0)}_{\mu \nu}(x')\;\;\;\;\;\;\;\;\;\;\;\;\;\text{spacelike region of P} (\mathbf{II}) \\
         &-\tilde{\phi}^{(0)}_{\mu \nu}(x')\;\;\;\;\;\;\;\;\;\;\text{past timelike region of P}(\mathbf{III})
     \end{cases}
\end{align}
Now we consider the function
\begin{align}
    F' = \lim_{\delta \rightarrow 0}\left(\frac{1}{2 z}(z^2 + |x - x'|^2 - (t' - t - i\epsilon)^2)\right)^{\delta}
\end{align}
This function vanishes when integrated against the boundary field $\phi^{(0)}_{\mu \nu}(x')$. It does not pick up any extra phase in the timelike region. Therefore, adding $F'$ to the normalizable kernel gives us the following kernel
\begin{align}
    \tilde{\mathbf{K}}_{N} = \mathbf{K}_{N} + b_{d}F'
\end{align}
This cancels out the kernel in the timelike region and gives a factor of $2$ in the spacelike region. The effective kernel is then given by
\begin{align}
    \tilde{\mathbf{K}}_{N} = \frac{\Gamma(\frac{d}{2})\Gamma(\frac{d}{2}+1)\Gamma(1 - \frac{d}{2})}{\pi^{\frac{d}{2} + 1}}\lim_{z' \rightarrow 0}\theta(\text{spacelike})\label{jejejeje1}
\end{align}
where we have used the value of $b_{d}$ and reintroduced the limit. Thus, we attain the spacelike kernels via the mode-sum approach.
\section{Green's function approach}\label{sec5}
We begin with the differential equations for the $p$- form and the graviton. The equations are written below as
\begin{align}
    z^{2}\partial^2_{z}A_{J} + (2 p - d + 1)z\partial_{z}A_{J} + z^2 \partial_{\alpha}\partial^{\alpha}A_{J} &= 0 \\
    \partial_\alpha \partial^\alpha h_{\mu \nu} + \partial^{2}_{z}h_{\mu \nu} + \frac{5-d}{z}\partial_{z}h_{\mu \nu} - \frac{2(d-2)}{z^2}h_{\mu \nu} &= 0
\end{align}
One does not expect either of the kernels obtained from these equations to be AdS covariant. That is because these wave equations cannot be cast solely in terms of the AdS covariant distance $\sigma$. However, it is possible to scale the fields by some $z^{\delta}$ so that the resulting equation of the new field is AdS covariant. We note that for the $p$- form, we can choose $\delta = - p$, and for the graviton, we choose $\delta = - 2$. The wave equations respectively turn out to be the following
\begin{align}
    z^2 \partial^2_{z}\Psi_{J} + z^{2}\partial_\alpha \partial^\alpha \Psi_{J} + (1 - d)z \partial_z \Psi + p (d - p)\Psi &= 0,\quad\quad\quad \Psi_{J} = z^{p}A_{J}\\
    z^2 \partial^2_{z}\Phi_{\mu \nu} + (1 - d)z\partial_{z} \Phi_{\mu \nu} + z^2 \partial_\alpha \partial^\alpha \Phi_{\mu \nu} &= 0,\quad\quad\quad \Phi_{\mu \nu} = z^2 h_{\mu \nu}
\end{align}
With this redefinition, the equation for $\Psi_{J}$ and $\Phi_{\mu \nu}$ can be shown to be equivalent to an ODE in terms of AdS chordal length $\sigma$. To do this, we consider Euclidean AdS and note the following relations
\begin{align}
    \frac{\partial \mathcal{C}}{\partial z} &= \Big(\frac{1}{z'} - \frac{\sigma}{z}\Big)\frac{\partial\mathcal{C}}{\partial \sigma} \\
    \frac{\partial^{2}\mathcal{C}}{\partial z^{2}} &= \Big( \frac{2 \sigma}{z^{2}} - \frac{1}{z z'} \Big)\frac{\partial \mathcal{C}}{\partial \sigma} + \Big( \frac{1}{z'} - \frac{\sigma}{z} \Big)^{2}\frac{\partial^{2}\mathcal{C}}{\partial \sigma^{2}}\\
    \frac{\partial^{2}\mathcal{C}}{\partial x^{2}} &= \Big( \frac{2 \sigma}{z z'} - \frac{1}{z^{2}} - \frac{1}{z'^{2} }\Big) \frac{\partial^{2}\mathcal{C}}{\partial \sigma^{2}} + \frac{d}{z z'}\frac{\partial \mathcal{C}}{\partial \sigma}
\end{align}
where $\mathcal{C}$ stands for either $\Psi_{J}$ or $\Phi_{\mu \nu}$. Using these relations, we find that the two equations reduce to 
\begin{align}
    (\sigma^2 - 1)\frac{d^2 \Psi_{J}}{d \sigma^2} + (1 + d)\sigma \frac{d \Psi_{J}}{d \sigma} + p (d - p)\Psi_{J} = 0 \\
    (\sigma^2 - 1)\frac{d^2 \Phi_{\mu \nu}}{d \sigma^2} + (1 + d)\sigma \frac{d \Phi_{\mu \nu}}{d \sigma} = 0
\end{align}
The most general solution to an equation of the form
\begin{align}
    (\sigma^2 - 1)G_{E}'' + (1 + d)\sigma G_{E}' + \Delta ( d - \Delta)G_{E} = 0
\end{align}
is given by
\begin{align}
    G_{E}(\sigma) = (\sigma^2 - 1)^{-\mu/2}\left( c_{1}\mathbf{P}^{\mu}_{\nu}(\sigma) + c_{2}\mathbf{Q}^{\mu}_{\nu}(\sigma)\right)
\end{align}
where $\mathbf{P}^{\mu}_{\nu}$ and $\mathbf{Q}^{\mu}_{\nu}$ are Legendre functions of type $3$ \cite{Heemskerk:2012mn}, with the constants $\mu = \frac{d - 1}{2}$ and $\nu = \Delta - \left(\frac{d + 1}{2}\right)$. To obtain the function on the spacelike cut, we demand that its real part vanishes in the timelike region. In that region, one obtains the following form for the \textit{Lorentzian} Green's function, defined via the analytic continuation $G_{M}(\sigma) \equiv i G_{E}(\sigma + i\epsilon)$
\begin{align}
    G_{M}(\sigma) = i c_{1}(-1)^{\mu}(1 - \sigma^2)^{-\mu/2}P^{\mu}_{\nu}(\sigma) + i c_{2} (-1)^{\mu}(1 - \sigma^2)^{-\mu/2}\left[Q^{\mu}_{\nu}(\sigma) - i \frac{\pi}{2}P^{\mu}_{\nu}(\sigma)\right]
\end{align}
This gives us the following relation for the coefficients $c_{1} = \frac{i \pi}{2}c_{2}$. The functions $P$ and $Q$ are the ordinary Legendre functions. Therefore, the function takes the following form in the spacelike region
\begin{align}
    \mathfrak{R}G_{M}(\sigma) = -c_{2}\left[\frac{\pi}{2}(\sigma^2 - 1)^{-\mu/2}\mathbf{P}^{\mu}_{\nu}(\sigma)\right]\theta(\text{spacelike})\label{normGreenFunc}
\end{align}
We ignore the divergent piece that arises from $\mathbf{Q}$ on the light-cone $\sigma = 1$ since that only plays a role when working with interactions (which we do not consider here). The value of $c_{2}$ is fixed via the short distance behavior of $G_{E}(\sigma)$ in the flat space limit. The result is independent of the tensor structure of the bulk field and depends only on the space-time dimensions. As an example, the same is discussed for the graviton in the following section, and the value \eqref{c2Val} is obtained. That is the value we shall substitute for $c_{2}$ in this section as well.  

The next step is, then, to insert this Green's function in Green's theorem to obtain the kernel for the $p$- form and the graviton (which corresponds to $\Delta = p$ and $\Delta = 0$ respectively)

To proceed with this, we shall first consider the $p$- form field.
\subsection{$p$- form}
We recall Green's function for the $p$- form normalizable mode \eqref{normGreenFunc}. In this case, we have the parameters $\mu = \frac{d - 1}{2}$ and $\nu = p - \frac{d}{2} - \frac{1}{2}$. The Green's theorem is given by the expression
\begin{align}
    \Psi_{J}(z, x) = \int \mathrm{d}^{d}x' \sqrt{\gamma'}\left(\Psi(z', x')\partial_{z'}G(\sigma) - G(\sigma)\partial_{z'}\Psi(z', x')\right)\Big\vert_{z' \rightarrow 0}\label{GreenTheorem}
\end{align}
With the appropriate Green's function, the theorem can be used to reconstruct any field (since it is an identity). We shall refer to the specific form \eqref{GreenTheorem} since we are interested in chordal Green's functions and the kernels obtained thereof. The final ingredient we require to proceed with the computation is the series expansion of the field $\Psi_{J}$. For this, we use \eqref{expansionPform} and use the scaling of $\Psi_{J}$ with respect to $A_{J}$. This gives us the following series 
\begin{align}
    \Psi_{J}(z, x) = \sum_{n = 0}^{\infty}z^{2 n - 2 \nu - 1 + p}\phi_{J}^{(n)}(x) + z^{2 n + p}j_{J}^{(n)}(x)\label{FUCK!}
\end{align}
Now we have to break the analysis into $2$ cases. These cases correspond to the range of $p$ with respect to $d$. We first begin with the case where $\nu > 0$. In terms of $p$ and $d$, this condition is $p > \frac{d + 1}{2}$ i.e. above the self-dual point (which we denote by $p_{0} = \frac{d - 1}{2}$). In this case, the $z' \rightarrow 0$ limit of the Green's function and its derivative are given by
\begin{align}
    G(\sigma)\vert_{z' \rightarrow 0} &= \frac{2^{-\nu}\sigma^{\nu - \mu}\Gamma(2 \nu + 1)}{\Gamma(1 + \nu)\Gamma(1 - \mu + \nu)} \label{GreenpLim1} \\
    \partial_{z'}G(\sigma)\vert_{z' \rightarrow 0} &= -\frac{2^{-\nu}\sigma^{\nu - \mu}\Gamma(2 \nu + 1)(\nu - \mu)}{\Gamma(1 + \nu)\Gamma(1 - \mu + \nu)z'} \label{GreenpLim2}
\end{align}
And correspondingly, the leading order contributions of the field $\Psi_{J}$ are given by
\begin{align}
    \Psi_{J}(z',x')\vert_{z' \rightarrow 0} &= z'^{p - 2\nu - 1}\phi_{J}^{(0)}(x') + z'^{p}j_{J}^{(0)}(x') \label{FieldpLim1}\\
    \partial_{z'}\Psi_{J}(z',x')\vert_{z' \rightarrow 0} &= (p - 2\nu - 1) z'^{p - 2\nu - 2}\phi_{J}^{(0)}(x') + p z'^{p-1}j_{J}^{(0)}(x') \label{FieldpLim2}
\end{align}
Plugging \eqref{GreenpLim1} - \eqref{FieldpLim2} into \eqref{GreenTheorem}, we find the following result
\begin{align}
    \Psi_{J}(z,x) &= \int \mathrm{d}^{d}x'\,z'^{-d + 1}\Big\{ -\left(z'^{p - 2\nu - 1}\phi_{J}^{(0)}(x') + z'^{p}j_{J}^{(0)}(x')\right)\frac{2^{-\nu}\sigma^{\nu - \mu}\Gamma(2 \nu + 1)(\nu - \mu)}{\Gamma(1 + \nu)\Gamma(1 - \mu + \nu)z'}  \notag\\
    &- \left((p - 2\nu - 1) z'^{p - 2\nu - 2}\phi_{J}^{(0)}(x') + p z'^{p-1}j_{J}^{(0)}(x')\right)\frac{2^{-\nu}\sigma^{\nu - \mu}\Gamma(2 \nu + 1)}{\Gamma(1 + \nu)\Gamma(1 - \mu + \nu)} \Big\}\Big\vert_{z' \rightarrow 0}
\end{align}
Combining the terms proportional to $\phi_{J}^{(0)}(x')$ and $j_{J}^{(0)}(x')$ respectively, we obtain
\begin{align}
\Psi_{J}(z,x) &= \int \mathrm{d}^{d}x'\Big\{-\frac{2^{-\nu}\Gamma(2 \nu + 1)(\nu - \mu + p)}{\Gamma(1 + \nu)\Gamma(1 - \mu + \nu)}\Big\}z'^{p - d}\sigma^{\nu - \mu}j_{J}^{(0)}(x')\Big\vert_{z' \rightarrow 0}\label{hehehe1}
\end{align}
From this, we can read off the kernel as follows (here $\pm$ in the index stands for the $\nu > 0$ and $\nu < 0$ regimes, respectively)
\begin{align}
    \mathbf{\tilde{K}}_{N, +}(z,x;x')\Big\vert_{p > p_{0}} = -c_{2}\frac{\pi}{2}\frac{2^{-\frac{d}{2}+p+\frac{1}{2}} \Gamma \left(-\frac{d}{2}+p+1\right)}{\sqrt{\pi } \Gamma (-d+p+1)}\lim_{z' \rightarrow 0}(\sigma z')^{p - d}\theta(\text{spacelike})
\end{align}
where the coefficient $c_{2}$ is introduced since the overall coefficient of the Green's function is not fixed yet. Also, the $\theta(\text{spacelike})$ is re-introduced for clarity. Re-inserting the value of $c_{2}$
\begin{align}
    \mathbf{\tilde{K}}_{N, +}(z,x;x')\Big\vert_{p > p_{0}} = \frac{(-1)^{\frac{d - 1}{2}}2^{p-d}\Gamma(p - \frac{d}{2} + 1)}{\pi^{\frac{d}{2}}\Gamma(p-d+1)}\lim_{z' \rightarrow 0}(\sigma z')^{p - d}\theta(\text{spacelike})\label{__hehehe1}
\end{align}
For the case of $\nu < 0$, the Green's function has the parameter $\nu$ replaced by $-\nu - 1$\footnote{This is due to the fact that $\mathbf{P}^{\mu}_{\nu}(x) = \mathbf{P}^{\mu}_{-\nu - 1}(x)$}. This implies the following limiting cases
\begin{align}
    G(\sigma)\vert_{z' \rightarrow 0} &= \frac{2^{\nu + 1}\sigma^{-\nu - \mu - 1}\Gamma(-2 \nu - 1)}{\Gamma(- \nu)\Gamma( - \mu - \nu)} \label{GreenpLim1.1} \\
    \partial_{z'}G(\sigma)\vert_{z' \rightarrow 0} &= -\frac{2^{\nu + 1}\sigma^{-\nu - \mu - 1}\Gamma(-2 \nu - 1)(-\nu - \mu - 1)}{\Gamma(-\nu)\Gamma(- \mu - \nu)z'} \label{GreenpLim2.1}
\end{align}
Plugging \eqref{GreenpLim1.1}-\eqref{GreenpLim2.1} and \eqref{FieldpLim1}-\eqref{FieldpLim2} into \eqref{GreenTheorem}, we obtain the following result.
\begin{align}
    \Psi_{J}(z,x) &= \int \mathrm{d}^{d}x'\,z'^{-d + 1}\Big\{ -\left(z'^{p - 2\nu - 1}\phi_{J}^{(0)}(x') + z'^{p}j_{J}^{(0)}(x')\right)\frac{2^{\nu + 1}\sigma^{-\nu - \mu - 1}\Gamma(-2 \nu - 1)(-\nu - \mu - 1)}{\Gamma(-\nu)\Gamma(- \mu - \nu)z'}\notag\\
    &- \left((p - 2\nu - 1) z'^{p - 2\nu - 2}\phi_{J}^{(0)}(x') + p z'^{p-1}j_{J}^{(0)}(x')\right)\frac{2^{\nu + 1}\sigma^{-\nu - \mu - 1}\Gamma(-2 \nu - 1)}{\Gamma(- \nu)\Gamma( - \mu - \nu)} \Big\}\Big\vert_{z' \rightarrow 0}
\end{align}
This expression reduces to 
\begin{align}
    \Psi_{J}(z,x) = -\int \mathrm{d}^{d}x'\,z'^{-d + 1}z'^{p - 2 \nu - 1 + d}\frac{2^{\nu + 2}(\frac{d}{2} - p)\sigma^{-\nu - \mu - 1}\Gamma(-2 \nu - 1)}{\Gamma(- \nu)\Gamma( - \mu - \nu)}\phi_{J}^{(0)}(x')\Big\vert_{z' \rightarrow 0}\label{hehehe2}
\end{align}
The kernel can be read from this
\begin{align}
    \mathbf{\tilde{K}}_{N, -}(z,x;x')\Big\vert_{p < p_{0}} = -c_{2}\frac{\pi}{2}\frac{2^{\frac{1}{2} (d-2 p-1)} \Gamma \left(\frac{d}{2}-p + 1\right)}{\sqrt{\pi } \Gamma (1-p)}\lim_{z' \rightarrow 0}(\sigma z')^{-p}\theta(\text{spacelike})
\end{align}
Re-inserting the value of $c_{2}$, we recover the following expression
\begin{align}
    \mathbf{\tilde{K}}_{N, -}(z,x;x')\Big\vert_{p < p_{0}} = \frac{(-1)^{\frac{d-1}{2}}2^{-p}\Gamma(\frac{d}{2} - p + 1)}{\pi^{\frac{d}{2}}\Gamma(1-p)}\lim_{z' \rightarrow 0}(\sigma z')^{-p}\theta(\text{spacelike})\label{__hehehe2}
\end{align}
We note that $\mathbf{\tilde{K}}_{N, \pm}$ are related to the original kernel $\mathbf{K}_{N, \pm}$ by a factor of $z^{-p}$.

Now, we consider the Non-normalisable mode and attempt to reconstruct the bulk field via the appropriate Green's function. It is not known what the exact procedure is for deriving this Green's function, but a natural choice is a solution complementary to \eqref{normGreenFunc}, restricted to the spacelike region. That solution is the following
\begin{align}
    \mathcal{G}_{M} = c_{3}(\sigma^2 - 1)^{-\mu/2}\mathbf{Q}^{\mu}_{\nu}(\sigma) =  \frac{2^{-p}C_{p}}{2 p - d}\sigma^{-p}\;_{2}F_{1}\left(\frac{p}{2}, \frac{1 + p}{2}; p - \frac{d}{2} + 1; \frac{1}{\sigma^2}\right)
\end{align}
where $\mathbf{Q}^{\mu}_{\nu}(\sigma)$ is the associated Legendre function of the $2^{\text{nd}}$ kind and is known to have the corresponding hypergeometric representation. The $\theta(\text{spacelike})$ is implied.

In order to fix the constant $C_{p}$, we follow the Lorentzian generalization of the Euclidean argument in \cite{Witten:1998qj}. We define note that in the $z' \rightarrow 0$ limit, $\mathcal{G}_{M}$ behaves as
\begin{align}
    \mathcal{G}_{M}\Big\vert_{z' \rightarrow 0} = \frac{2^{-p}C_{p}}{2 p - d}\left(\frac{2 z' z}{z^2 + z'^2 + \vert x - x' \vert^2}\right)^{p} \equiv \frac{z'^{p}}{2 p - d}K_{p}(z, x; x')
\end{align}
The function $K_{p}(z, x; x')$ is the bulk-boundary propagator, which is not the same (but is related to) the kernel. This propagator should have a $\delta$ function normalization, which we impose via the following integral
\begin{align}
    \int_{\text{spacelike}} \mathrm{d}^{d}x\, K_{p}(z, x; x') = -C_{p}z^{d-p}\pi^{\frac{d}{2} - 1}\Gamma(1 - p)\Gamma(p - \frac{d}{2})\cos\pi p 
\end{align}
Setting this integral to $1$ fixes the coefficient $C_{p} = -\frac{2^{-p}\Gamma(p)\tan\pi p}{\Gamma( p -\frac{d}{2})\pi^{\frac{d}{2}}}$. Using this value of $C_{p}$, the non-normalizable Green's function is given by
\begin{align}
    \mathcal{G}_{M} = -\frac{2^{-p-1}\Gamma(p)\tan\pi p}{\Gamma( p -\frac{d}{2} + 1)\pi^{\frac{d}{2}}}\sigma^{-p}\;_{2}F_{1}\left(\frac{p}{2}, \frac{ p  + 1}{2}; p -\frac{d}{2} + 1;\frac{1}{\sigma^2}\right)\theta(\text{spacelike})
\end{align}
\noindent
The leading order behaviour for $\mathcal{G}_{M}(\sigma)$ and $\partial_{z'}\mathcal{G}_{M}(\sigma)$ at $z' \rightarrow 0$ is given by
\begin{align}
    \mathcal{G}_{M}(\sigma)\Big\vert_{z' \rightarrow 0} &= -\frac{2^{-p-1}\Gamma(p)\tan\pi p}{\Gamma( p -\frac{d}{2} + 1)\pi^{\frac{d}{2}}}\sigma^{-p}\\
    \partial_{z'}\mathcal{G}_{M}(\sigma)\Big\vert_{z' \rightarrow 0} &= -\frac{2^{-p-1}\Gamma(p + 1)\tan\pi p}{\Gamma( p -\frac{d}{2} + 1)\pi^{\frac{d}{2}}z'}\sigma^{-p}
\end{align}
Plugging this in the Green's theorem with \eqref{FieldpLim1}-\eqref{FieldpLim2} gives us
\begin{align}
    \Psi_{J}(z,x) &= \int \mathrm{d}^{d}x'\,z'^{-d + 1}\Big\{ -\left(z'^{p - 2\nu - 1}\phi_{J}^{(0)}(x') + z'^{p}j_{J}^{(0)}(x')\right)\frac{2^{-p-1}\Gamma(p + 1)\tan\pi p}{\Gamma( p -\frac{d}{2} + 1)\pi^{\frac{d}{2}} z'}\sigma^{-p}\notag\\
    &+ \left((p - 2\nu - 1) z'^{p - 2\nu - 2}\phi_{J}^{(0)}(x') + p z'^{p-1}j_{J}^{(0)}(x')\right)\frac{2^{-p-1}\Gamma(p)\tan\pi p}{\Gamma( p -\frac{d}{2} + 1)\pi^{\frac{d}{2}}}\sigma^{-p}\Big\}\Big\vert_{z' \rightarrow 0}
\end{align}
These terms combine in an appropriate fashion to give us the following expression
\begin{align}
    \Psi_{J}(z,x) &= -\int \mathrm{d}^{d}x'\,z'^{-p}\frac{2^{-p-1}\Gamma(p)\tan\pi p}{\Gamma( p -\frac{d}{2} + 1)\pi^{\frac{d}{2}} }(2 p - d)\sigma^{-p}\phi_{J}^{(0)}(x')\Big\vert_{z'\rightarrow 0}\label{hahaha1}
\end{align}
From where we can read off the kernel as follows
\begin{align}
    \mathbf{\tilde{K}}_{nN, +} (z, x; x') = -\frac{2^{-p}\Gamma(p)\tan\pi p}{\Gamma( p -\frac{d}{2})\pi^{\frac{d}{2}}}\lim_{z' \rightarrow 0}(\sigma z')^{-p}\theta(\text{spacelike})
\end{align}
Using identities involving the gamma function, one can show that the kernel is equivalent to
\begin{align}
    \mathbf{\tilde{K}}_{nN, +} (z, x; x') = \frac{(-1)^{\frac{d-1}{2}}\Gamma(1 + \frac{d}{2} - p)2^{-p}}{\pi^{\frac{d}{2}}\Gamma(1 - p)}\lim_{z' \rightarrow 0}(\sigma z')^{-p}\theta(\text{spacelike})\label{__hahaha1}
\end{align}
Now, we consider the other case where $\nu < 0$. This corresponds to replacing the $\mathbf{Q}^{\mu}_{\nu}$ with $\mathbf{Q}^{\mu}_{-\nu - 1}$. This is valid operation since $\mathbf{Q}^{\mu}_{-\nu - 1}$ is the solution that is complementary to $\mathbf{P}^{\mu}_{-\nu - 1}$. The Green's function is written as
\begin{align}
    \mathcal{G}_{M} = c'_{3}(\sigma^2 - 1)^{-\mu/2}\mathbf{Q}^{\mu}_{-\nu - 1}(\sigma) = -\frac{2^{p - d}C'_{p}}{2 p -  d}C'_{p}\sigma^{p - d}\;_{2}F_{1}\left(\frac{d - p}{2}, \frac{d - p + 1}{2}; 1 + \frac{d}{2} - p;\frac{1}{\sigma^2} \right)
\end{align}
Normalising this function in an identical manner as the $\nu > 0$ case gives us $C'_{p} = -\frac{\Gamma(d - p)\tan\pi(d - p)}{\pi^\frac{d}{2}\Gamma(\frac{d}{2} - p)}$. The full expression of the Greens' function becomes
\begin{align}
    \mathcal{G}_{M} = -\frac{2^{p - d - 1}\Gamma(d - p)\tan\pi(d - p)}{\pi^{\frac{d}{2}}\Gamma(\frac{d}{2} - p + 1)}\sigma^{p - d}\;_{2}F_{1}\left(\frac{d - p}{2}, \frac{d - p + 1}{2}; 1 + \frac{d}{2} - p;\frac{1}{\sigma^2} \right)\theta(\text{spacelike})
\end{align}
Using this, we write the limiting expression of $G(\sigma)$ and its' derivative as follows
\begin{align}
    \mathcal{G}_{M}(\sigma)\Big\vert_{z' \rightarrow 0} &= -\frac{2^{p - d - 1}\Gamma(d - p)\tan\pi(d - p)}{\pi^{\frac{d}{2}}\Gamma(\frac{d}{2} - p + 1)}\sigma^{p - d} \\
    \partial_{z'}\mathcal{G}_{M}(\sigma)\Big\vert_{z' \rightarrow 0} &= -\frac{2^{p - d - 1}\Gamma(d - p + 1)\tan\pi(d - p)}{\pi^{\frac{d}{2}}\Gamma(\frac{d}{2} - p + 1)z'}\sigma^{p - d}
\end{align}
Using these expressions along with \eqref{FieldpLim1}-\eqref{FieldpLim2} we obtain the result
\begin{align}
    &\Psi_{J}(z,x) = \int \mathrm{d}^{d}x' z'^{-d + 1}\Big\{ -\left(z'^{p - 2\nu - 1}\phi_{J}^{(0)}(x') + z'^{p}j_{J}^{(0)}(x')\right)\frac{2^{p - d - 1}\Gamma(d - p + 1)\tan\pi(d - p)}{\pi^{\frac{d}{2}}\Gamma(\frac{d}{2} - p + 1)z'}\sigma^{p - d}\notag\\
    &+ \left((p - 2\nu - 1) z'^{p - 2\nu - 2}\phi_{J}^{(0)}(x') + p z'^{p-1}j_{J}^{(0)}(x')\right)\frac{2^{p - d - 1}\Gamma(d - p)\tan\pi(d - p)}{\pi^{\frac{d}{2}}\Gamma(\frac{d}{2} - p + 1)}\sigma^{p - d}\sigma^{\nu - \mu}\Big\}\Big\vert_{z' \rightarrow 0}
\end{align}
This expression simplifies to the following
\begin{align}
    \Psi_{J}(z,x) = \int d^{d}x'(2 p  - d)\frac{2^{p - d - 1}\Gamma(d - p)\tan\pi(d - p)}{\pi^{\frac{d}{2}}\Gamma(\frac{d}{2} - p + 1)}(\sigma z')^{p - d}j_{J}^{(0)}(x')\Big\vert_{z' \rightarrow 0}\label{hahaha2}
\end{align}
From this, we can read off the kernel as follows
\begin{align}
    \mathbf{\tilde{K}}_{n N, -}(z, x; x') = -\frac{2^{p - d}\Gamma(d - p)\tan\pi(d - p)}{\pi^{\frac{d}{2}}\Gamma(\frac{d}{2} - p)}\lim_{z' \rightarrow 0}(\sigma z')^{p - d}\theta(\text{spacelike})
\end{align}
It can be seen that, via some properties of gamma functions, this expression reduces to
\begin{align}
    \mathbf{\tilde{K}}_{n N, -}(z, x; x') =\frac{(-1)^{\frac{d-1}{2}}\Gamma(1  - \frac{d}{2} + p)2^{p - d}}{\pi^{\frac{d}{2}}\Gamma(1 - d + p)}\lim_{z' \rightarrow 0}(\sigma z')^{p - d}\theta(\text{spacelike})\label{__hahaha2}
\end{align}
Therefore we recover the AdS-covariant piece in the mode sum kernel via the Greens' function approach as well.

So far, we have dealt with either $\nu < 0$ or $\nu > 0$. Now we turn to the special case of $\nu = 0$, which is at $p_0 + 1$. The special feature of the $\nu = 0$ point is that the $z' \rightarrow 0$ limit of the two Green's functions differ by a factor of $\sigma$. 

The bulk solution's $z-$ dependence can be obtained by setting $\nu = 0$ in \eqref{FUCK!}. The two Green's functions, respectively have the following leading order behavior
\begin{align}
    G(\sigma)\Big\vert_{z' \rightarrow 0} &= \frac{(-1)^{\frac{d + 1}{2}}}{2^{\frac{d + 1}{2}}\pi^{\frac{d - 1}{2}}}\frac{\sigma ^{-\mu }}{\Gamma (1-\mu )}\theta(\text{spacelike}) \\
    \partial_{z'}G(\sigma)\Big\vert_{z' \rightarrow 0} &= \frac{(-1)^{\frac{d + 1}{2}}\mu}{2^{\frac{d + 1}{2}}\pi^{\frac{d - 1}{2}}}\frac{\sigma ^{-\mu }}{z' \Gamma (1-\mu )}\theta(\text{spacelike}) \\
    \mathcal{G}_{M}(\sigma)\Big\vert_{z' \rightarrow 0} &= -\frac{2^{-p-1}\Gamma(p)\tan\pi p}{\Gamma( p -\frac{d}{2} + 1)\pi^{\frac{d}{2}}}\sigma^{-p}\theta(\text{spacelike}) \\
    \partial_{z'}\mathcal{G}_{M}(\sigma)\Big\vert_{z' \rightarrow 0} &= -\frac{2^{-p-1}\Gamma(p + 1)\tan\pi p}{z'\Gamma( p -\frac{d}{2} + 1)\pi^{\frac{d}{2}}}\sigma^{-p}\theta(\text{spacelike})
\end{align}
Plugging the Green's function for the normalisable mode $G(\sigma)$ into Green's theorem along with \eqref{FUCK!} gives the following expression (by using $p = \mu + 1$)
\begin{align}
    \Psi_{J}(z, x; x') &= \frac{(-1)^{\frac{d + 1}{2}}}{2^{\frac{d + 1}{2}}\pi^{\frac{d - 1}{2}}}\int \mathrm{d}^{d}x'\,z'^{-d + 1}\Big\{ \left(z'^{\mu}\phi_{J}^{(0)}(x') + z'^{\mu + 1}j_{J}^{(0)}(x')\right)\mu \frac{\sigma ^{-\mu }}{z' \Gamma (1-\mu )} \notag\\
    &-\left(\mu z'^{\mu - 1}\phi_{J}^{(0)}(x')+ (\mu + 1) z'^{\mu}j_{J}^{(0)}(x')\right)\frac{\sigma ^{-\mu }}{\Gamma (1-\mu )}
    \Big\}\Big\vert_{z' \rightarrow 0}
\end{align}
This expression simplifies to 
\begin{align}
    \Psi_{J}(z, x; x') = -\frac{(-1)^{\frac{d + 1}{2}}}{2^{\frac{d + 1}{2}}\pi^{\frac{d - 1}{2}}}\int \mathrm{d}^{d}\vec{x'}z'^{-d + 1}\Big\{ z'^{\mu}\frac{\sigma^{-\mu}}{\Gamma(1 - \mu)}j_{J, 0}(x')
    \Big\}\Big\vert_{z' \rightarrow 0}\label{hohoho1}
\end{align}
From this, the kernel can be read off as follows
\begin{align}
    \mathbf{\tilde{K}}_{N, 0} = -\frac{(-1)^{\frac{d - 1}{2}}}{2^{\frac{d + 1}{2}}\pi^{\frac{d - 1}{2}}\Gamma(\frac{3 - d}{2})}\lim_{z' \rightarrow 0}(\sigma z')^{\frac{1 - d}{2}}\theta(\text{spacelike})
\end{align}
where the subscript $0$ is to indicate the $\nu = 0$ point. Turning to the Non-normalisable mode, we find the following expression by inserting the Green's function $\mathcal{G}_{M}(\sigma)$ into Green's theorem and using \eqref{FUCK!}.
\begin{align}
    \Psi_{J}(z, x; x') &= \int \mathrm{d}^{d}x'\,z'^{-d + 1}\Big\{ -\left(z'^{\mu}\phi_{J}^{(0)}(x') + z'^{\mu + 1}j_{J}^{(0)}(x')\right)\frac{2^{-p-1}\Gamma(p + 1)\tan\pi p}{z'\Gamma( p -\frac{d}{2} + 1)\pi^{\frac{d}{2}}}\sigma^{-p} \notag\\
    &+\left(\mu z'^{\mu - 1}\phi_{J}^{(0)}(x') + (\mu + 1) z'^{\mu}j_{J}^{(0)}(x')\right)\frac{2^{-p-1}\Gamma(p)\tan\pi p}{\Gamma( p -\frac{d}{2} + 1)\pi^{\frac{d}{2}}}\sigma^{-p}
    \Big\}\Big\vert_{z' \rightarrow 0}
\end{align}
This expression simplifies to
\begin{align}
    \Psi_{J}(z, x; x') = \int \mathrm{d}^{d}x'\,z'^{-d + 1}\Big\{ z'^{\mu - 1}\frac{2^{-p-1}\Gamma(p + 1)\tan\pi p}{z'\Gamma( p -\frac{d}{2} + 1)\pi^{\frac{d}{2}}}\sigma^{-p}\phi_{J}^{(0)}(x')
    \Big\}\Big\vert_{z' \rightarrow 0}\label{hohoho2}
\end{align}
From this, we can read off the kernel
\begin{align}
    \mathbf{\tilde{K}}_{n N, 0} = \frac{i^{d+1} (2 \pi )^{\frac{1}{2}-\frac{d}{2}}}{(d+1) \Gamma \left(-\frac{d}{2}-\frac{1}{2}\right)}\lim_{z' \rightarrow 0}(\sigma z')^{-\frac{d + 1}{2}}\theta(\text{spacelike})
\end{align}
And so we derive the expressions for the mode kernels at the $\nu = 0$ point.
\subsection{Graviton}

To study the Green's function approach for the graviton, we consider the series expansion in $z$ 
\begin{align}
    \Phi_{\mu \nu}(z,x) &=  \sum_{k = 0}^{\infty}z^{d + 2 k}\phi^{(k)}_{\mu \nu}(x) + \sum_{k=0}^{\infty}z^{2 k}j^{(k)}_{\mu \nu}(x)\label{expansionGrav}
\end{align}
The solution of the differential equation is 
\begin{align}
    \Phi_{\mu \nu}(\sigma) = c_1 +  c_2 (\sigma^2 - 1)^{-\mu/2}\mathbf{Q}^{\mu}_{\mu}(\sigma) 
\end{align}
Again, following the arguments of \cite{Heemskerk:2012mn} we find that the Green's function can be made spacelike for $c_{1} = -\frac{\pi c_{2}}{2}(2\mu - 1)!!$. Therefore, we note that the Green's function for the non-normalizable mode will simply turn out to be
\begin{align}
    G_{M}(\sigma) = -\frac{\pi c_{2}}{2}(2\mu - 1)!!\theta(\text{spacelike})
\end{align}
Using this expression and \eqref{expansionGrav} 
\label{sec:photon} in Green's theorem, we obtain the following result
\begin{align}
    \Phi_{\mu \nu}(z,x) = \int \mathrm{d}^{d}x'\,z'^{-d + 1}\Big\{ - c_{1}\left(d \, z'^{d - 1}\phi^{(0)}_{\mu \nu}(x')\right)\Big\}
\end{align}
From which the kernel can be read off as 
\begin{align}
    \mathbf{K}_{N}(z, x; x') = \frac{\pi c_{2} d}{2}(2\mu - 1)!!\theta(\text{spacelike})
\end{align}
Inserting the value of $c_{2}$ we find that
\begin{align}
    \mathbf{K}_{N}(z, x; x') = (-1)^{\frac{d+1}{2}} \frac{\Gamma \left(\frac{d}{2}+1\right)}{\pi ^{\frac{d}{2}}} \theta(\text{spacelike})\label{jajajaja1}
\end{align}
Now, we turn to the non-normalizable mode. As in the previous sections, we pick the following Green's function 
\begin{align}
    \mathcal{G}_{M}(\sigma) = c_2 (\sigma^2 - 1)^{-\mu/2}\mathbf{Q}^{\mu}_{\mu}(\sigma) = -\frac{2^{-d-1}\Gamma(d)\tan\pi d}{\Gamma(\frac{d}{2} + 1)\pi^\frac{d}{2}}\sigma^{-d}\;_{2}F_{1}\left(\frac{d}{2}, \frac{d + 1}{2}; \frac{d}{2} + 1; \frac{1}{\sigma^2}\right)
\end{align}
And in the $z' \rightarrow 0$ limit, the Green's function and its derivative takes the form
\begin{align}
    \mathcal{G}_{M}(\sigma)\Big\vert_{z' \rightarrow 0} &= -\frac{2^{-d-1}\Gamma(d)\tan\pi d}{\Gamma(\frac{d}{2} + 1)\pi^\frac{d}{2}}\sigma^{-d} \\
    \partial_{z'}\mathcal{G}_{M}(\sigma)\Big\vert_{z' \rightarrow 0} &= -\frac{2^{-d-1}\Gamma(d + 1)\tan\pi d}{\Gamma(\frac{d}{2} + 1)\pi^\frac{d}{2} z'}\sigma^{-d}
\end{align}
Plugging this into the Green's theorem with \eqref{expansionGrav}, we find the expression
\begin{align}
    \Phi_{\mu \nu}(z,x) &= c_{2}\int \mathrm{d}^{d}x'\, z'^{-d + 1}\Big\{ 
    -\left(z'^{d - 1}\phi^{(0)}_{\mu \nu}(x') + z'^{-1}j^{(0)}_{\mu \nu}(x') \right)\frac{2^{-d-1}\Gamma(d + 1)\tan\pi d}{\Gamma(\frac{d}{2} + 1)\pi^\frac{d}{2} }\sigma^{-d} \notag\\
    &+ d\,z'^{d - 1}\phi^{(0)}_{\mu \nu}(x')\frac{2^{-d-1}\Gamma(d)\tan\pi d}{\Gamma(\frac{d}{2} + 1)\pi^\frac{d}{2}}\sigma^{-d}
    \Big\}\Big\vert_{z' \rightarrow 0}
\end{align}
Upon simplification, this reduces to the following expression
\begin{align}
    \Phi_{\mu \nu}(z,x) =  -\int \mathrm{d}^{d}x'\frac{2^{-d-1}\Gamma(d + 1)\tan\pi d}{\Gamma(\frac{d}{2} + 1)\pi^\frac{d}{2} z'}(\sigma z')^{-d}\Big\vert_{z' \rightarrow 0}j^{(0)}_{\mu \nu}x'
\end{align}
And from this, the kernel can be read off as follows
\begin{align}
    \mathbf{K}_{n N}(z,x;x') = -\frac{2^{-d-1}\Gamma(d + 1)\tan\pi d}{\Gamma(\frac{d}{2} + 1)\pi^\frac{d}{2} z'}\lim_{z' \rightarrow 0}(\sigma z')^{-d}\theta(\text{spacelike})\label{jajajaja2}
\end{align}
And thus, we derive the expressions for the bulk reconstruction kernels for the two modes up to a constant factor. The constant can be determined from the short-distance behavior of Euclidean Green's function. Note that we require the transverse traceless modes at the boundary, which is given by
\begin{align}
    \langle h_{\mu\nu}(r) h_{\rho\sigma}(0)&=\mathcal{I}_{\mu\nu\rho\sigma}G_E(r).
\end{align}
Here $G_E(r)$ is the Euclidean scalar Green's function, and $\mathcal{I}_{\mu\nu\rho\sigma}$ obeys the following properties.
\begin{align}
\mathcal{I}_{\mu\nu\rho\sigma}&=\mathcal{I}_{\rho\sigma\mu\nu}\nonumber\\
 \mathcal{I}^{\mu}_{~\mu\rho\sigma}&=\mathcal{I}_{\mu\nu\rho}^{~~~\rho}=0\nonumber\\
\partial^{\mu}\mathcal{I}_{\mu\nu\rho\sigma}&=\partial^{\nu}\mathcal{I}_{\mu\nu\rho\sigma}=\partial^{\rho}\mathcal{I}_{\mu\nu\rho\sigma}=\partial^{\sigma}\mathcal{I}_{\mu\nu\rho\sigma}=0.
\end{align}
It is obvious that the tensor structure $\mathcal{I}_{\mu\nu\rho\sigma}$ enforces to have transverse and traceless degrees of freedom at the boundary. The short-distance behavior of the Euclidean scalar Green's function is well known
\begin{align}
   \lim_{r\rightarrow 0} G_{E}(r)\sim-\frac{1}{(d-1)\rm{vol}(S^d)r^{d-1}}.
\end{align}
Here $r$ is  the Euclidean radial coordinate  and $\rm{vol}(S^d)=\frac{\pi^{\frac{d+1}{2}}}{\Gamma(\frac{d+1}{2})}.$ At a short-distance, AdS chordal length becomes $\sigma\sim 1+\frac{r^2}{2 R^2}.$ Therefore, the Euclidean scalar Green's function is determined with the appropriate delta function source at the origin, and the constant $c_2$ is evaluated \cite{Hamilton:2006az}
\begin{align}
    c_2=\frac{(-1)^{\mu+1}}{2^{\mu-1}(d-1)\rm{vol}(S^d)\Gamma(\mu)R^{d-1}} = \frac{(-1)^{\frac{d + 1}{2}}}{2^{\frac{d - 1}{2}}\pi^{\frac{d + 1}{2}}}.\label{c2Val}
\end{align}
\label{sec:graviton}
Note that the constant is entirely determined from the scalar Green's function, and the tensor transformation properties of the graviton are encoded in $\mathcal{I}_{\mu\nu\rho\sigma}.$

\section{Summary of the main results}
At this point, it is useful to summarise the results of the $p$- form fields. So far, we have used the terms “normalizable” and “non-normalizable” as placeholders without really referring to their interpretations. The normalizable and non-normalizable modes are identified via the fall-off behavior of the holographic coordinate ($z$ in Poincar\'e coordinates) near the boundary $z = 0$. We recall that the two modes in mode expansion of the bulk $p$- form scale as \eqref{expansionPform}
\begin{align}
    A_{J}(z,x) \sim z^{d - 2 p}\phi^{(0)}_{J}(x) + j^{(0)}_{J}(x)
\end{align}
Above the self-dual point, we have $p > \frac{d + 1}{2}$. In this regime, the \textit{normalisable} mode is identified as $j^{(0)}_{J}$ while the \textit{non-normalisable} mode is identified as $\phi^{(0)}_{J}(x)$. This is due to the fact that the fall-off exponent for $z$ is a negative integer $< -1$. This behavior is also picked up by the normalizable \textit{Green’s function}, as is seen from \eqref{hehehe1} and \eqref{hahaha1} respectively. Similarly, below the self-dual point, we have $p < \frac{d + 1}{2}$. In this regime, the normalisable mode is identified as $\phi^{(0)}_{J}(x)$ and the non-normalisable mode as $j^{(0)}_{J}(x)$. Again, this is reflected in the Green’s function approach via the equations \eqref{hehehe2} and \eqref{hahaha2}, respectively. 

The final case is that of the self-dual point. At this point, we have $p = \frac{d + 1}{2}$. Therefore, the fall-off behavior of the mode corresponding to $\phi^{(0)}_{J}(x)$ is $z^{-1}$, which makes it non-normalizable and consequently the normalizable mode is $j^{(0)}_{J}(x)$. The same is once again reflected by the Green’s function, as can be seen from \eqref{hohoho1} and \eqref{hohoho2}, respectively. Due to the scenario being different above and below the self-dual point, it is convenient to represent the result in a concise table below. 
\begin{center}
    \begin{tabular}{||c|c|c|c|c||}
     \hline
     $p$- value & Normalisable mode & Kernel ($\mathbf{K}_{N}$) & Non-Normalisable mode & Kernel ($\mathbf{K}_{n N})$ \\[0.5ex]
     \hline\hline
     $p > p_{0}$ & $j^{(0)}_{J}(x)$ & Eqn \eqref{__hehehe1} \& \eqref{KnNpFormMDSUM} & $\phi^{(0)}_{J}(x)$ &  Eqn \eqref{__hahaha1} \& \eqref{KNpFormMDSUM}  \\[1ex]
     \hline
     $p \leq p_{0}$ & $\phi^{(0)}_{J}(x)$ & Eqn \eqref{__hehehe2} \& \eqref{KNpFormMDSUM} & $j^{(0)}_{J}(x)$ &  Eqn \eqref{__hahaha2} \& \eqref{KnNpFormMDSUM}\\[1ex]
     \hline
\end{tabular}
\end{center}
Here we have indicated the equations where the respective (Ads-covariant) kernel expressions are written via the Green’s function and mode sum approaches. Naturally, both approaches agree. In the first column we have replaced $p \geq \frac{d + 1}{2}$ by $p > p_0$ and $p < \frac{d + 1}{2}$ by $p \leq p_0$ where $p_0 = \frac{d - 1}{2}$, i.e. the self-dual point. The rationale behind doing so is the fact that $d$ is an odd integer, and so $\frac{d + 1}{2}$ and $\frac{d - 1}{2}$ are both integers (as is $p$) separated by $1$.

Turning to the graviton, we recall the following mode expansion of the bulk graviton field near $z = 0$
\begin{align}
    h_{\mu \nu}(z, x) \sim z^{d - 2}\phi^{(0)}_{\mu \nu}(x) + z^{-2}j^{(0)}_{\mu \nu}(x)
\end{align}
It is evident from the fall-off behavior of the two modes (except for $d = 1$, which corresponds to AdS$_{2}$ which we do not consider) that the normalizable mode is the one corresponding to $\phi^{(0)}_{\mu \nu}(x)$ and the non-normalizable mode is $j^{(0)}_{\mu \nu}(x)$. The same is reflected by the Green’s function approach, where the normalizable Green’s function picks $\phi^{(0)}_{\mu \nu}$ and vice versa. Naturally, the normalisable mode kernels obtained via both methods (equations \eqref{jajajaja1} and \eqref{jejejeje1}) match, as do the non-normalisable mode kernels (equations \eqref{jajajaja2} and \eqref{jejejeje2}).

Therefore, our results obtained via the mode sum and Green’s function methods agree for the two modes of the $p$- form and graviton fields.
\section{Conclusions}
In this paper, we focus our attention on deriving space-like kernels for $p$-forms and the graviton. We derive these kernels using two different approaches- mode-sum and Green’s function method. We show that these two approaches lead to the same space-like kernels for both modes. We also study the properties of these kernels under Hodge-dual transformation, and we find a mismatch between a kernel and its Hodge-dual. A similar mismatch has also been found in the partition function of $p$-forms \cite{Raj:2016zjp, David:2021wrw}. Since the duality holds true at the classical level, it is not obvious that a similar statement will also work for the smearing functions. However, it will be interesting to reproduce the mismatch of free energies under the Hodge-dual transformation from the smearing functions.

In this work, we restrict ourselves to the Poincar\'e patch of AdS. We keep the derivations of the kernels in the global coordinate for future work where we wish to establish the connections between kernels. It is important to note that the Poincar\'e
kernels we obtained via mode sums have a natural  AdS covariant form, up to some extra terms. We wish to come up with a general argument for dropping those extra terms. In this regard, $i\epsilon$ prescription given in \cite{Bhattacharjee:2022ehq} will be useful.
\label{sec:conclusions}

It will be interesting to develop the HKLL procedure for higher derivative conformal fields and conformal higher spin fields. These theories provide essential tools to study conformal field theories in higher dimensions, including free energies and conformal anomalies \cite{Beccaria:2014jxa, Beccaria:2017dmw, Mukherjee:2021alj, Mukherjee:2021rri}. However, these theories are non-unitary in general due to the presence of higher derivatives in the kinetic term in the action. But it will be interesting to extract the boundary data using a bulk reconstruction procedure, in particular, for Weyl graviton and conformal higher derivative gauge theories. It will be interesting to reproduce the boundary two-point functions and the central charge for these non-unitary conformal field theories from the bulk. 

It will also be interesting to develop the HKLL reconstruction procedure where the scalar and fermionic field is coupled to the gauge field and gravity. A similar kind of questions have already been addressed in \cite{Heemskerk:2012np}. It will be nice to obtain the smearing functions by solving for Green’s functions for the spatial equations and applying Green’s theorem. Similarly, one can also obtain the smearing functions using the mode-sum approach and see the equivalence between these two methods. It is important to have a general statement about the smearing functions when the matter is coupled to gravity. Although there is significantly more gauge redundancy in the case of gravity, one can perform a perturbative analysis to have a perturbative explanation of holography.

Lastly, the underlying motivation for evaluating the HKLL kernels for these fields has been to develop a better understanding of the nature of solutions of wave equations. We expect this to teach us about bulk reconstruction in other geometries. One of the future goals would then be to extend this prescription to other geometries relevant to holography, such as Minkowski or de-Sitter spacetimes (for example, \cite{BHOWMICK2019134977}). 

\acknowledgments

We thank Chethan Krishnan and Debajyoti Sarkar for valuable discussions and comments regarding this manuscript. The authors acknowledge the hospitality of the International Center For Theoretical Physics (ICTP) for the duration of the Spring School on Superstring Theory, during which part of this work was completed. BB is partially supported by the Ministry of Human Resource Development, Govt. of India through the Prime Ministers' Research Fellowship. 
\appendix
\section{Derivation of bulk equation of $p$-forms in Poincar\'e coordinates}\label{appA}
In this appendix, we derive the bulk equation of free $p$-forms in the Poincar\'e patch.
The covariant equation of motion of $p$-forms is given by
\begin{align}
    \nabla_{M}F^{M,J_{1},\dots,J_{p}} = 0\label{nablaF}
\end{align}
The field strength tensor $F^{M J_{1},\dots,J_{p}}$ can be explicitly written in terms of the  completely antisymmetric gauge potential
\begin{align}
    F^{M,J_{1},\dots,J_{p}} &= \nabla^{M}A^{J_{1},\dots,J_{p}} + (-1)^{p}\nabla^{J_{1}}A^{J_{2},\dots,J_{p},M} + \cdots \notag\\ &\cdots + (-1)^{p k}\nabla^{J_{k}}A^{J_{k+1},\dots,J_{p}, M, J_{1},\dots, J_{p-1}} + \cdots + (-1)^{p^2}\nabla^{J_{p}}A^{M, J_{1},\dots,J_{p-1}}\label{pformfieldstrength}
\end{align}
The first term of \eqref{nablaF} is given by $\nabla_{M}\nabla^{M}A^{J_{1},\dots,J_{p}}$. We evaluate this expression later. For now, we begin by focusing on the general term $\nabla_{M}\nabla^{J_{k}}A^{J_{k+1},\dots,J_{p}, M, J_{1},\dots, J_{k-1}}$. To simplify this term, we consider the following identity for a general tensor $T^{A_{1},\dots,A_{n}}$
\begin{align}
    [\nabla_{M},\nabla_{N}]T^{A_{1},\dots,A_{n}} = \sum_{i = 1}^{n}R^{A_{i}}_{K M N}T^{A_{1},\dots,A_{i-1}, K, A_{i+1},\dots, A_{n}}\label{reimann}
\end{align}
where $R^{A}_{KMN}$ is the Reimann tensor.

From the covariant gauge condition $\nabla_{J_1}A^{J_1,\dots,J_p} = 0$ and \eqref{reimann}, we find that

\begin{align}
    \nabla_{M}\nabla^{J_1}A^{J_2,\dots,J_{p},M} &= g^{J_{1} N}\nabla_{M}\nabla_{N}A^{J_{2},\dots,J_{p},M} \notag\\
    &= g^{J_{1} N}\left(R^{J_2}_{K M N}A^{K, J_{3},\dots,J_{p}, M} + \cdots + R^{J_{p}}_{K M N}A^{J_{2},\dots,J_{p-1},K,M} + R^{M}_{K M N}A^{J_{2},\dots,J_{p},K}\right) \notag\\
    &= - d A^{J_{2},\dots,J_{p},J_{1}} - A^{J_{1},J_{3},\dots,J_{p},J_{2}} -\cdots- A^{J_{2},\dots,J_{p-1},J_{1},J_{p}}
\end{align}
Using the antisymmetric nature of the $p$- form field, we can write this expression as 
\begin{align}
    \nabla_{M}\nabla^{J_1}A^{J_2,\dots,J_{p},M} = (-1)^{p}(d-(p - 1))A^{J_{1},\dots,J_{p}}
\end{align}
It is straightforward to see that each term subsequent term in \eqref{pformfieldstrength}, when acted upon by $\nabla_{M}$, picks up additional factors of $(-1)^{p}$. Thus, the equation of motion takes the simple form
\begin{align}
    \nabla_{M}\nabla^{M}A^{J_{1},\dots,J_{p}} + p(d- p + 1)A^{J_{1},\dots,J_{p}} = 0
\end{align}
To evaluate this expression, we need to evaluate $g^{M N}\nabla_{M}\nabla_{N}A^{J_{1},\dots,J_{p}}$. For this purpose, it is easier to introduce a new notation:
\begin{align}
    A^{J} &\equiv A^{J_{1},\dots,J_{p}} \\
    A^{J,\{J_{i}, K\}} &\equiv A^{J_{1},\dots,J_{i-1},K,J_{i + 1},\dots,J_{p}} \\
    A^{J,\{J_{i}, K\},\{J_{j}, Q\}} &\equiv A^{J_{1},\dots,J_{i-1},K,J_{i + 1},\dots,J_{j-1},Q,J_{j+1},\dots,J_{p}}
\end{align}
In this notation, we can write the term $\nabla_{M}\nabla_{N}A^{J}$ as follows
\begin{align}
    \nabla_{M}\nabla_{N}A^{J} &= \partial_{M}\partial_{N}A^{J} + \sum_{i}\Gamma^{J_{i}}_{M K}\partial_{N}A^{J,\{J_{i}, K\}} - \Gamma^{K}_{M N}\partial_{K}A^{J} \notag\\
    &+ \sum_{i}\partial_{M}\left(\Gamma^{J_{i}}_{N K}A^{J,\{J_{i}, K\}}\right) + \sum_{i}\sum_{j}\Gamma^{J_{j}}_{M Q}\Gamma^{J_{i}}_{N K}A^{J,\{J_{i}, K\},\{J_{j}, Q\}} - \sum_{i}\Gamma^{Q}_{M N}\Gamma^{J_{i}}_{Q K}A^{J, \{J_{i}, Q \}}\label{fulleqnpform}
\end{align}
Plugging in the Christoffel symbols for empty AdS spacetime, 
\begin{align}
    \Gamma^{A}_{z B} &= -\frac{1}{z}\delta^{A}_{B} \\
    \Gamma^{z}_{\mu \nu} &= \frac{1}{z}\eta_{\mu \nu}
\end{align}
The equation \eqref{fulleqnpform} reduces to (that is, $g^{M N}\nabla_{M}\nabla_{N}A^{J} + p(d- p + 1)A^{J} = 0$)
\begin{align}
    \partial_{N}\partial^{N}A^{J} - (d - 1 + 2 p)z\partial_{z}A^{J} + 2 d p A^{J} = 0
\end{align}
which, splitting in $z$ and boundary coordinates, becomes
\begin{align}
    z^{2}\partial^2_{z}A^{J} + z^{2}\partial_{\alpha}\partial^{\alpha}A^{J} - (d - 1 + 2 p)z\partial_{z}A^{J} + 2 d p A^{J} = 0
\end{align}
For the covariant $p$- form, i.e. for the field $A_{J_{1},\dots,J_{p}} \equiv A_{J}$, the equation of motion becomes
\begin{align}
    z^{2}\partial^2_{z}A_{J} + (2 p - d + 1)z\partial_{z}A_{J} + z^2 \partial_{\alpha}\partial^{\alpha}A_{J} = 0
\end{align}

\section{Massive $p$-forms} 
\label{AppMassP}
In this section, we present HKLL bulk reconstruction kernels of massive $p$-forms in the  Poincar\'e patch of even $AdS_{d+1}$. The bulk equation is given by
\begin{align}
     z^{2}\partial^2_{z}A_{J} + (2 p - d + 1)z\partial_{z}A_{J} + z^2 \partial_{\alpha}\partial^{\alpha}A_{J} -m^2A_J= 0\label{deqmassp}
\end{align}
The solution to this bulk equation is obtained as
\begin{align}\label{massp}
 A_{J}(z, x) = \int_{q \geq 0}\frac{\mathrm{d}^{d}q}{(2 \pi)^{d}}z^{\frac{1}{2} (d-2 p)}\left(c_{J}(q) J_{\frac{1}{2} \sqrt{(d-2 p)^2+4 m^2}}(|q| z)+ d_{J}(q) Y_{\frac{1}{2} \sqrt{(d-2 p)^2+4 m^2}}(q z)\right)e^{i q. x}.
 \end{align}
 We use the mode sum approach to obtain the kernels for $p$-forms. Given the bulk solution in \eqref{massp}, we can have the asymptotic expansion of the solution around $z=0$
 \begin{align}
    A_{J}(x,z)&=z^{-p}\sum_{n=0}^{\infty}\left(z^{\Delta+2n}\phi_{J}^{(2 n)}(x)+z^{d-\Delta+2n}j_{J}^{(2 n)}(x)
    \right)\nonumber\\
    &=  A^{N}_{J}(z, x)+ A^{n N}_{J}(z, x) , 
\end{align}
where $\Delta=\frac{d}{2}+ \sqrt{(\frac{d}{2}- p)^2+ m^2}$ and two modes are given by
\begin{align}
    A^{N}_{J}(z, x) = \sum_{n = 0}^{\infty} z^{2 n+\Delta-p}\phi_{J}^{(2 n)}(x) ,\quad\quad
    A^{n N}_{J}(z, x) = \sum_{n = 0}^{\infty} z^{2 n + d - p - \Delta}j_{J}^{(2 n)}(x) \label{expansionmassPform}
\end{align}
This consideration holds for $\Delta - \frac{d}{2} \notin\text{Integers}$. For the scenario of $\Delta - \frac{d}{2} \in \text{Integers}$, the analysis will closely follow Appendix.\ref{appB}.

The coefficients $\phi_{J}^{(2 n)}(x)$ and $j_{J}^{(2 n)}(x)$ can be extracted from the bulk solutions and these serve as boundary data. Following the massless $p$-form fields, we choose coefficients corresponding to $n=0$ case as data. We now follow the similar steps from \eqref{kerp} to \eqref{nonNormEvenAdSkernel}, we obtain the kernels for massive $p$-forms in even $AdS$
\begin{align}
    \begin{split}
         \mathbf{K}_{N}(z, x; x')
    = \frac{z^{\Delta -p} \Gamma(d/2)}{2 \pi^{\frac{d}{2} + 1}X^{d}}\;_{2}F_{1}\left(\frac{d}{2}, 1;1-\frac{d}{2}+\Delta; - \frac{z^2}{X^2}\right)\\
    \mathbf{K}_{nN}(z, x; x') = \frac{z^{d - \Delta - p}\Gamma(d/2)}{2 \pi^{d/2 + 1}X^{d}}\;_{2}F_{1}\left(\frac{d}{2}, 1; 1 - \Delta + \frac{d}{2}; - \frac{z^2}{X^2}\right)
    \end{split}
\end{align}
Note that, the functional form of the kernels depends on the conformal dimension $\Delta$ and in the massless limit one recovers the kernels \eqref{EvenAdSkernelp1} and \eqref{EvenAdSNNkernelp1}. The identification of the kernels as corresponding to normalizable or non-normalizable modes depends on the exponents in series \eqref{expansionmassPform}. For $\Delta \geq -1$, the $\phi^{(0)}_{J}(x)$ mode is identified as normalizable and vice-versa. 

Using the hypergeometric identity \eqref{hypIden}, one can rewrite the kernels as follows
\begin{align}
    z^{p}\mathbf{K}_{N}(z, x; x') &= \frac{\Gamma(1 - \frac{d}{2} + \Delta)\Gamma(\frac{d}{2} - 1)}{2\pi^{\frac{d}{2} + 1}\Gamma(\Delta - \frac{d}{2})}\frac{z^{\Delta - 2}}{X^{d-2}}\;_{2}F_{1}\left(1, 1 + \frac{d}{2} - \Delta; 2 - \frac{d}{2}; - \frac{X^2}{z^2}\right) \notag \\ 
    &+ \frac{2^{\Delta - d}\Gamma(1 - \frac{d}{2} + \Delta)(-1)^{\frac{d-1}{2}}}{2\pi^{\frac{d}{2} + 1}\Gamma(1 - d + \Delta)}\lim_{z' \rightarrow 0}(\sigma z')^{\Delta - d}\label{MassPFEqN} \\
    z^{p}\mathbf{K}_{n N}(z, x; x') &= \frac{\Gamma(1 - \Delta + \frac{d}{2})\Gamma(\frac{d}{2} - 1)}{2\pi^{\frac{d}{2} + 1}\Gamma(\frac{d}{2} - \Delta)}\frac{z^{d - \Delta - 2}}{X^{d - 2}}\;_{2}F_{1}\left(1, 1 + \Delta - \frac{d}{2}; 2 - \frac{d}{2}; - \frac{X^2}{z^2}\right) \notag \\
    &+ \frac{2^{-\Delta}\Gamma(1 - \Delta + \frac{d}{2})(-1)^{\frac{d-1}{2}}}{2\pi^{\frac{d}{2} + 1}\Gamma(1 - \Delta)}\lim_{z' \rightarrow 0}(\sigma z')^{-\Delta}\label{MassPFEqnN}
\end{align}
From these expressions, one can note that the first terms in \eqref{MassPFEqN} and \eqref{MassPFEqnN} do not contain the correct powers of $z$, when compared to the series \eqref{expansionmassPform}. The second terms are respectively the AdS-covariant pieces and so after utilising a prescription to drop the first terms, the second terms can be written in the following spacelike form (after an antipodal mapping)
\begin{align}
    z^{p}\mathbf{K}_{N}(z, x; x') &= \frac{2^{\Delta - d}\Gamma(1 - \frac{d}{2} + \Delta)(-1)^{\frac{d-1}{2}}}{\pi^{\frac{d}{2} + 1}\Gamma(1 - d + \Delta)}\lim_{z' \rightarrow 0}(\sigma z')^{\Delta - d}\theta(\text{spacelike}) \label{zackN}\\
    z^{p}\mathbf{K}_{n N}(z, x; x') &= \frac{2^{-\Delta}\Gamma(1 - \Delta + \frac{d}{2})(-1)^{\frac{d-1}{2}}}{\pi^{\frac{d}{2} + 1}\Gamma(1 - \Delta)}\lim_{z' \rightarrow 0}(\sigma z')^{-\Delta}\theta(\text{spacelike})\label{zacknN}
\end{align}
Turning to the Green's function approach, we note that the equation \eqref{deqmassp} can be cast in the following form
\begin{align}
    (\sigma^2 - 1)\frac{d^2}{d\sigma^2}\Psi_{J}(\sigma) + (d  + 1)\sigma \frac{d}{d\sigma}\Psi_{J}(\sigma) - \Delta(\Delta - d)\Psi_{J}(\sigma) = 0
\end{align}
where $\sigma$ is the chordal distance and $\Psi_{J} = z^{p}A_{J}$. The solution of this equation has the form
\begin{align}
    \Psi_{J}(\sigma) = c_{1}(\sigma^2 - 1)^{-\mu/2}\mathbf{P}^{\mu}_{\nu}(\sigma) + c_{2}(\sigma^2 - 1)^{-\mu/2}\mathbf{Q}^{\mu}_{\nu}(\sigma)
\end{align}
where $\mathbf{P}^{\mu}_{\nu}$ and $\mathbf{Q}^{\mu}_{\nu}$ stand for the associated Legendre polynomials of the third kind. The parameters are $\mu = \frac{d - 1}{2}$ and $\nu = \Delta - \frac{d + 1}{2}$. From this solution, the normalizable mode Green's function can be readily evaluated similar to the massless case. These are again different for $\nu > 0$ and $\nu < 0$.
\begin{align}
    G_{M}(\sigma) = -c_{2}\frac{\pi}{2}(\sigma^2 - 1)^{-\mu/2}\mathbf{P}^{\mu}_{\nu}(\sigma)\theta(\text{spacelike})
\end{align}
for $\nu > 0$, with $c_{2}$ derived in \eqref{c2Val}. For $\nu < 0$, the only change is that the index $\nu$ in the above equation is replaced by $-\nu - 1$. 

Using this Green's function in Green's theorem, one recovers the kernels \eqref{zackN} and \eqref{zacknN} for $\nu < 0$ and $\nu > 0$ respectively. The Green's function for the non-normalizable mode can be derived in a similar way as the massless case. The explicit expression is the following (for $\nu > 0$)
\begin{align}
    \mathcal{G}_{M}(\sigma) =  - \frac{2^{-\Delta - 1}\Gamma(\Delta)\tan\pi\Delta}{\Gamma(\Delta - \frac{d}{2} + 1)\pi^{\frac{d}{2}}}\sigma^{-\Delta}\;_{2}F_{1}\left(\frac{\Delta}{2}, \frac{\Delta + 1}{2}; \Delta - \frac{d}{2} + 1; \frac{1}{\sigma^2}\right)\theta(\text{spacelike})
\end{align}
For the $\nu < 0$, the replacement is $\Delta \rightarrow d - \Delta$. Once again, using this function in Green's theorem gives \eqref{zacknN} and \eqref{zackN} for $\nu < 0$ and $\nu > 0$ respectively. With this, we conclude this brief section on the massive extension of the massless $p$-form discussed in depth in the main text. The results for both cases mirror each other appropriately.

Returning to the expression for $\Delta$, we find that the Breitenlohner-Freedman (BF) bound for the massive $p-$ form to be $\Delta > \frac{d}{2}$ \cite{David:2020mls, Witten:1998qj}. It is important to note that the kernels can be formally extended beyond the BF bound. It has been achieved via appropriate analytic continuations in \cite{DelGrosso:2019gow, Aoki:2021ekk}.  An similar approach may be used to extend the kernels derived here beyond the BF bound, such that we consider the window $\frac{d}{2} - 1 \leq \Delta \leq \frac{d}{2} + 1$. We direct the reader to a more detailed discussion regarding the same in \cite{Bhattacharjee:2022ehq}. Within such a BF window, the normalizable and non-normalizable modes of the scaled $p$-form $\Psi_{J}$ can be identified with CFT operators (for the non-normalizable mode, in the Legendre transformed CFT). 
\section{Mode sum kernels in odd AdS$_{d + 1}$ : $p$-form}\label{appB}
The odd AdS case is a significantly more complicated one. The solution for the bulk wave equations is the following
\begin{align}
    A_{J}(z, x) = \int_{q \geq 0}z^{\nu}\left(a_{J}(q)J_{\nu}(q z) + b_{J}(q)Y_{\nu}(q z)\right)e^{i q. x}\frac{\mathrm{d}^{d}q}{(2 \pi)^{d}}\label{OddAdSsoln}
\end{align}
This part has to be handled carefully since the combination $Y_{\nu}$ contains a $J_{\nu}$ piece as well. Denoting $\nu = n$, and using the series representation of $Y_{n}$, we have
\begin{align}
    Y_{n}(q z) = \frac{2}{\pi}\ln\left(\frac{q z}{2}\right)J_{n}(q z) - \frac{(q z)^{n}}{2^{n}\pi}\sum_{k = 0}^{\infty}\alpha_{k, n}(q^2 z^2)^{k} - \frac{(q z)^{-n}}{2^{-n}\pi}\sum_{k = 0}^{n - 1}\beta_{k, n}(q^2 z^2)^{k}
\end{align}
where $\alpha_{k, n} = \frac{(-1)^{k}}{4^{k}\Gamma(k + 1)\Gamma(k + n + 1)}(\psi(k + 1) + \psi(k + n + 1))$ and $\beta_{k, n} = \frac{\Gamma(n - k)}{\Gamma( k + 1)4^{k}}$. Taking these terms, along with the series form of $J_{n}$ into account, we find the following expression for $A_{J}(z, x)$
\begin{align}
    A_{J}(z, x) = \sum_{k = 0}^{\infty}z^{2 n + 2 k}\phi_{k}(x) + \sum_{k = 0}^{\infty}\ln(z) z^{2 n + 2 k}\tilde{\phi}_{k}(x) + \sum_{k = 0}^{n - 1}z^{2 k}j_{k}(x)
\end{align}
where
\begin{align}
    j_{k}(x) &= \int_{q \geq 0}b_{J}\beta_{k, n}2^{n}q^{2 k - n}e^{i q. x}\frac{\mathrm{d}^{d}q}{(2\pi)^{d}} \\
    \tilde{\phi}_{k}(x) &= \int_{q \geq 0}\frac{2}{\pi}b_{J}\frac{(-1)^{k}q^{n + 2 k}}{2^{n}4^{k}\Gamma(k + 1)\Gamma(k + n + 1)}e^{i q. x}\frac{\mathrm{d}^{d}q}{(2\pi)^{d}} \\
    \phi_{k}(x) &= \int_{q \geq 0}a_{J}\frac{(-1)^{k}q^{n + 2 k}}{2^{n}4^{k}\Gamma(k + 1)\Gamma(k + n + 1)}e^{i q. x}\frac{\mathrm{d}^{d}q}{(2\pi)^{d}} \notag \\ &+ \int_{q \geq 0}b_{J}\left(\frac{2}{\pi}\ln(\frac{q}{2})\frac{(-1)^{k}q^{n + 2 k}}{2^{n}4^{k}\Gamma(k + 1)\Gamma(k + n + 1)} - \frac{\alpha_{k, n}}{2^{n}}q^{n + 2 k}\right)e^{i q. x}\frac{\mathrm{d}^{d}q}{(2\pi)^{d}}
\end{align}
These results should be inverted to obtain the expression for $b_{J}$ and $a_{J}$. There is a choice to make here regarding the boundary data. For this purpose, we choose the two independent boundary pieces to be $j_{0}$ and $\phi_{0}$. This is the natural choice since it corresponds to the $z^{2 k}$ and $z^{2 k - 2 n}$ fall-offs, analogous to the even AdS case. 

Inverting the expressions for $j_{k}(x)$ gives us
\begin{align}
    b_{J}(q) = \frac{1}{\beta_{k, n}}\int q^{n - 2 k}2^{-n}j_{k}(x')e^{-i q. x'}\mathrm{d}^{d}x'
\end{align}
Inserting this in the expression for $\phi_{k}$ and using it to invert the relation to find $a_{J}$, we get
\begin{align}
    &\int \phi_{k}(x')e^{-i q. x'}\mathrm{d}^{d}x' - b_{J}\left(\frac{2}{\pi}\ln(\frac{q}{2})\frac{(-1)^{k}q^{n + 2 k}}{2^{n}4^{k}\Gamma(k + 1)\Gamma(k + n + 1)} - \frac{\alpha_{k, n}}{2^{n}}q^{n + 2 k}\right) = a_{J}\frac{(-1)^{k}q^{n + 2 k}}{2^{n}4^{k} k! \Gamma(k + n + 1)}
\end{align}
From this, we can extract the expression for $a_{J}(q)$
\begin{align}
    a_{J}(q) &= \frac{2^{n}4^{k}\Gamma(k + 1)\Gamma(k + n + 1)}{(-1)^{k}q^{n + 2 k}}\int \phi_{k}(x')e^{-i q. x'}\mathrm{d}^{d}x' - \frac{2}{\pi}\ln(\frac{q}{2})\frac{1}{\beta_{k, n}}\int q^{n - 2 k}2^{-n}j_{k}(x')e^{-i q. x'}\mathrm{d}^{d}x'\notag \\ & + \alpha_{k, n}(-1)^{k}\Gamma(k + 1)\Gamma(k + n + 1)4^{k}\frac{1}{\beta_{k, n}}\int q^{n - 2 k}2^{-n}j_{k}(x')e^{-i q. x'}\mathrm{d}^{d}x'
\end{align}
These expressions are messy. But we can make it simpler by choosing $k = 0$, since any two pieces of boundary data are equivalent. So, we have the following expressions
\begin{align}
    b_{J}(q) &= \frac{q^{n}}{2^{n}\Gamma(n)}\int j_{0}(x')e^{-i q. x'}\mathrm{d}^{d}x' \\
    a_{J}(q) &= \frac{2^{n}\Gamma(n + 1)}{q^{n}}\int \phi_{0}(x')e^{-i q. x'}\mathrm{d}^{d}x' - \left( \frac{2 \ln(q/2)}{\pi}  + \gamma - \psi(n + 1)\right)\frac{q^{n}}{2^{n}\Gamma(n)}\int j_{0}(x')e^{-i q. x'}\mathrm{d}^{d}x' 
\end{align}
Thankfully, these are much simpler expressions that we can begin to evaluate carefully by inserting back into \eqref{OddAdSsoln}
\begin{align}
    A_{J}(z, x) &= z^{n}\int \frac{2^{n}\Gamma(n + 1)}{q^{n}}J_{n}(q z)e^{i q. (x - x')}\phi_{0}(x')\frac{\mathrm{d}^{d}q \mathrm{d}^{d}x'}{(2\pi)^{d}}  + z^{n}\int \frac{q^{n}}{2^{n}\Gamma(n)}Y_{n}(q z)e^{i q. (x - x')}j_{0}(x')\frac{\mathrm{d}^{d}q \mathrm{d}^{d}x'}{(2\pi)^{d}} \notag\\
    &- z^{n}\int \left( \frac{2 \ln(q/2)}{\pi}  + \gamma - \psi(n + 1)\right)\frac{q^{n}}{2^{n}\Gamma(n)}J_{n}(q z)e^{i q. (x - x')}j_{0}(x')\frac{\mathrm{d}^{d}q \mathrm{d}^{d}x'}{(2\pi)^{d}}
\end{align}
From this, we read off the two kernel integrals. We denote them by $\mathbf{K}_{N}$ for $\phi_{0}$ and $\mathbf{K}_{n N}$ for $j_{0}$, analogous to the even AdS case. This gives us the following integrals
\begin{align}
    \mathbf{K}_{N}(z,x; x') &= z^{n}2^{n}\Gamma(n + 1)\int_{q \geq 0}q^{-n}J_{n}(q z)e^{i q.(x - x')}\frac{\mathrm{d}^{d}q}{(2\pi)^{d}} \label{NormOddAdSkernel} \\
    \mathbf{K}_{n N}(z, x; x') &= \frac{z^{n}}{2^{n}\Gamma(n)}\int_{q \geq 0}q^{n}Y_{n}(q z)e^{i q. (x - x')}\frac{\mathrm{d}^{d}q}{(2\pi)^{d}}\notag\\ &- \frac{z^{n}}{2^{n}\Gamma(n)}\int_{q \geq 0} \left( \frac{2 \ln(q/2)}{\pi}  + \gamma - \psi(n + 1)\right)q^{n}J_{n}(q z)e^{i q. (x - x')}\frac{\mathrm{d}^{d}q}{(2\pi)^{d}} \label{nonNormOddAdSkernel}
\end{align}
\subsection{Evaluating the integrals}
This, right here, is the Herculean task: evaluating these integrals. Evaluating the first integral \eqref{NormOddAdSkernel} is simple. This integral was evaluated for the even AdS case as well \eqref{NormEvenAdSkernel}. Therefore, the result \eqref{EvenAdSkernelp1} can be used directly, with the identification $n = \frac{d}{2} - p$. Thus, we have
\begin{align}
    \mathbf{K}_{N}(z, x; x') = \frac{z^{2 n}\Gamma(d/2)}{2 \pi^{d/2 + 1}X^{d}}\;_{2}F_{1}\left(\frac{d}{2}, 1; n + 1; - \frac{z^2}{X^2}\right)
\end{align}
which reduces to \eqref{EvenAdSkernelp1} by putting in the expression for $n$ in terms of $\Delta$ and $d$. 

Evaluating the kernel $\mathbf{K}_{n N}$ is a whole different beast altogether. We begin with the seemingly simpler integral
\begin{align}
    I_{1} = \int_{q \geq 0}q^{n}Y_{n}(q z)e^{i q. (x - x')}\frac{\mathrm{d}^{d}q}{(2\pi)^{d}}
\end{align}
which is the first half of \eqref{nonNormOddAdSkernel}. Using \eqref{identity_integral-0}, this integral reduces to 
\begin{align}
    I_{1} = \frac{1}{\pi (2 \pi)^{d/2}X^{d/2 - 1}}\int_{x = 0}^{\infty}x^{n + d/2}Y_{n}(x z)K_{\frac{d}{2} - 1}(x X)\mathrm{d}x
\end{align}
The final integral that we have to evaluate in order to determine $I_{1}$ is
\begin{align}
    I_{2} = \int_{x = 0}^{\infty}x^{n + d/2}Y_{n}(x z)K_{\frac{d}{2} - 1}(x X)\mathrm{d}x
\end{align}
To evaluate it, we utilize the following relation
\begin{align}
    Y_{n}(x z) = -\frac{i^{n}}{\pi}\left((-1)^{n}K_{n}(-i x z) + K_{n}(i x z) \right)
\end{align}
This gives us the following type of integrals
\begin{align}
    \int_{x = 0}^{\infty}x^{n + d/2}K_{n}(\pm i x z)K_{\frac{d}{2}-1}(x X)\mathrm{d}x
\end{align} 
To evaluate this, we turn to the following identity
\begin{align}
    \int_{x = 0}^{\infty}x^{-\lambda}K_{\mu}(a x) K_{\nu}(b x)\mathrm{d}x &= \frac{2^{-2 - \lambda}a^{-\nu + \lambda - 1}b^{\nu}}{\Gamma(1 - \lambda)}\Gamma(\frac{1 - \lambda + \mu + \nu}{2})\Gamma(\frac{1 - \lambda - \mu + \nu}{2})\Gamma(\frac{1 - \lambda + \mu - \nu}{2})\notag \\
    &\times\Gamma(\frac{1 - \lambda - \mu - \nu}{2}) \;_{2}F_{1}(\frac{1 - \lambda + \mu + \nu}{2}, \frac{1 - \lambda - \mu + \nu}{2}; 1 -\lambda; 1 - \frac{b^2}{a^2})
\end{align}
with the constraints $\text{Re}(a + b) > 0,\;\;\text{Re}(\lambda) < 1 - \vert \text{Re}(\mu)\vert - \vert \text{Re}(\nu) \vert$. All of these conditions are satisfied for both the integrals that have to be evaluated. Therefore, we have
\begin{align}
    \int_{x = 0}^{\infty}x^{-\lambda}K_{\mu}(a x) K_{\nu}(b x)\mathrm{d}x = \frac{2^{-2 + n + \frac{d}{2}}X^{\frac{d}{2} - 1}(\pm i z)^{-n - d}}{n + \frac{d}{2}}\Gamma(d/2)\Gamma(n + 1)\;_{2}F_{1}\left(n + \frac{d}{2}, \frac{d}{2}; 1 + n + \frac{d}{2}; 1 + \frac{X^{2}}{z^2}\right)
\end{align}
Thus, the integral $I_{2}$ becomes
\begin{align}
    I_{2} = -\frac{2^{-1 + n + \frac{d}{2}}i^{-d}X^{\frac{d}{2} - 1}(z)^{-n - d}}{\pi(n + \frac{d}{2})}\Gamma(d/2)\Gamma(n + 1)\;_{2}F_{1}\left(n + \frac{d}{2}, \frac{d}{2}; 1 + n + \frac{d}{2}; 1 + \frac{X^{2}}{z^2}\right)
\end{align}
From this, we can write $I_{1}$ as
\begin{align}
    I_{1} = -\frac{2^{-1 + n}i^{-d}z^{-n - d}}{ \pi^{\frac{d}{2} + 2}(n + \frac{d}{2})}\Gamma(d/2)\Gamma(n + 1)\;_{2}F_{1}\left(n + \frac{d}{2}, \frac{d}{2}; 1 + n + \frac{d}{2}; 1 + \frac{X^{2}}{z^2}\right)
\end{align}
Thus the first term of the kernel \eqref{nonNormOddAdSkernel} (which we denote at $T_{1}$) becomes
\begin{align}
    T_{1} = -\frac{(-1)^{d/2}z^{-2 n - d}n\Gamma(d/2)}{ 2 \pi^{\frac{d}{2} + 2}(n + \frac{d}{2})}\;_{2}F_{1}\left(n + \frac{d}{2}, \frac{d}{2}; 1 + n + \frac{d}{2}; 1 + \frac{X^{2}}{z^2}\right)\label{T1expr}
\end{align}
The next term of \eqref{nonNormOddAdSkernel} is quite the non-trivial one. For this, we consider the integral
\begin{align}
    I_{3} = \int_{q \geq 0}\ln(q) q^{n}J_{n}(q z)e^{i q. (x - x')}\frac{\mathrm{d}^{d}q}{(2\pi)^{d}}
\end{align}
This integral, upon using \eqref{identity_integral-0}, reduces to the following effective integral
\begin{align}
    I_{4} =  \int_{x = 0}^{\infty}\ln(x)x^{n + d/2}J_{n}(x z)K_{\frac{d}{2}-1}(x X)\mathrm{d}x 
\end{align}

One way is to use the following parametrization of the logarithm
\begin{align}
    \ln(x) = \lim_{\epsilon \rightarrow 0}\frac{x^{\epsilon} - 1}{\epsilon}
\end{align}
This splits $I_{4}$ into two parts
\begin{align}
    I_{4, 1} &= \int_{x = 0}^{\infty}x^{n + d/2 + \epsilon}J_{n}(x z)K_{\frac{d}{2}-1}(x X)\mathrm{d}x \notag\\
    I_{4, 2} &= \int_{x = 0}^{\infty}x^{n + d/2}J_{n}(x z)K_{\frac{d}{2}-1}(x X)\mathrm{d}x \notag
\end{align}
For the first part, the integral becomes the following, using \eqref{identity_integral-1}
\begin{align}
    I_{4, 1} &= \frac{z^{n}\Gamma(n + \frac{d}{2} + \frac{\epsilon}{2})\Gamma(n + 1 + \frac{\epsilon}{2})}{2^{-n - \frac{d}{2} - \epsilon + 1} X^{2 n + \frac{d}{2} +\epsilon + 1}\Gamma(1 + n)}\;_{2}F_{1}\left(n + \frac{d}{2} + \frac{\epsilon}{2}, n + 1 + \frac{\epsilon}{2}; 1 + n; -\frac{z^{2}}{X^{2}} \right)\\
    I_{4, 2} &= \frac{z^{n}\Gamma(n + \frac{d}{2})\Gamma(n + 1)}{2^{-n - \frac{d}{2} + 1} X^{2 n + \frac{d}{2} + 1}\Gamma(1 + n)}\;_{2}F_{1}\left(n + \frac{d}{2}, n + 1; 1 + n; -\frac{z^{2}}{X^{2}} \right)
\end{align}
Therefore, we can write the integral $I_{4}$ as follows
\begin{align}
    I_{4} &= \lim_{\epsilon \rightarrow 0}\frac{z^{n}\Gamma(n + \frac{d}{2} + \frac{\epsilon}{2})\Gamma(n + 1 + \frac{\epsilon}{2})}{2^{-n - \frac{d}{2} - \epsilon + 1} X^{2 n + \frac{d}{2} +\epsilon + 1}\Gamma(1 + n)\epsilon}\;_{2}F_{1}\left(n + \frac{d}{2} + \frac{\epsilon}{2}, n + 1 + \frac{\epsilon}{2}; 1 + n; -\frac{z^{2}}{X^{2}} \right) \notag \\
    &-\frac{z^{n}\Gamma(n + \frac{d}{2})\Gamma(n + 1)}{2^{-n - \frac{d}{2} + 1} X^{2 n + \frac{d}{2} + 1}\Gamma(1 + n)\epsilon}\;_{2}F_{1}\left(n + \frac{d}{2}, n + 1; 1 + n; -\frac{z^{2}}{X^{2}} \right)
\end{align}
This evaluates to the following result
\begin{align}
    I_{4} &= 2^{\frac{d}{2}+n-2} z^n X^{-\frac{d}{2}-2 n-1} \Gamma \left(\frac{d}{2}+n\right) \left(\frac{z^2}{X^2}+1\right)^{-\frac{d}{2}-n} \left(\psi\left(\frac{d}{2}+n\right)+\psi(n+1)-2 \log (X)+\log (4)\right) \notag \\
    &+ 2^{\frac{d}{2}+n-2} z^n X^{-\frac{d}{2}-2 n-1} \Gamma \left(\frac{d}{2}+n\right) \left(G^{1}\left(n+1,\frac{d}{2}+n,n+1,-\frac{z^2}{X^2}\right)+G^{2}\left(n+1,\frac{d}{2}+n,n+1,-\frac{z^2}{X^2}\right)\right)
\end{align}
where $G^{1}(a, b, c) = \frac{\partial}{\partial x}\;_{2}F_{1}(x, b; c; z)\Big\vert_{x = a}$ and $G^{2}(a, b, c) = \frac{\partial}{\partial x}\;_{2}F_{1}(a, x; c; z)\Big\vert_{x = b}$. 

Thus we obtain the second term of the non-normalizable integral as follows
\begin{align}
    T_{2} &= - \frac{z^{-n}}{2^{n}\Gamma(n)}\int_{q \geq 0}\frac{2 \ln(q)}{\pi}q^{n}J_{n}(q z)e^{i q. (x - x')}\frac{\mathrm{d}^{d}q}{(2\pi)^{d}} \notag\\
    &= -\frac{\Gamma \left(\frac{d}{2}+n\right)}{2 \pi^{\frac{d}{2} + 2}\Gamma(n)} \left(z^{2} + X^{2} \right)^{-\frac{d}{2}-n} \left(\psi\left(\frac{d}{2}+n\right)+\psi(n+1)-2 \log (X)+\log (4)\right) \notag \\
    &- \frac{ X^{-d - 2 n-1} \Gamma \left(\frac{d}{2}+n\right)}{2 \pi^{\frac{d}{2} + 2}\Gamma(n)} \left(G^{1}\left(n+1,\frac{d}{2}+n,n+1,-\frac{z^2}{X^2}\right)+G^{2}\left(n+1,\frac{d}{2}+n,n+1,-\frac{z^2}{X^2}\right)\right)\label{T2expr}
\end{align}
The final remaining term is quite straightforward. The integral we need to evaluate is
\begin{align}
    \int_{q \geq 0} q^{n}J_{n}(q z)e^{i q. (x - x')}\frac{\mathrm{d}^{d}q}{(2\pi)^{d}} &= \frac{1}{\pi (2\pi)^{d/2}X^{\frac{d}{2} - 1}}\int_{x = 0}^{\infty}x^{n + d/2}J_{n}(x z)K_{\frac{d}{2} - 1}(x X)\mathrm{d}x \notag \\
    &= \frac{z^{n}\Gamma(n + \frac{d}{2})\Gamma(n + 1)}{2^{-n + 1} \pi^{\frac{d}{2} + 1} X^{2 n + d}\Gamma(1 + n)}\;_{2}F_{1}\left(n + \frac{d}{2}, n + 1; 1 + n; -\frac{z^{2}}{X^{2}} \right)
\end{align}
Thus we have the final term in the kernel
\begin{align}
    T_{3} = \left(\psi(n + 1) + \frac{2 \ln 2}{\pi} - \gamma \right)\frac{\Gamma(n + \frac{d}{2})}{2 \pi^{\frac{d}{2} + 1} X^{2 n + d}\Gamma(n)}\;_{2}F_{1}\left(n + \frac{d}{2}, n + 1; 1 + n; -\frac{z^{2}}{X^{2}} \right)
\end{align}
Using the fact that $\;_{2}F_{1}(a,b;b,z) = (1-z)^{-a}$, we have
\begin{align}
    T_{3} = \left(\psi(n + 1) + \frac{2 \ln 2}{\pi} - \gamma \right)\frac{\Gamma(n + \frac{d}{2})}{2 \pi^{\frac{d}{2} + 1}\Gamma(n)}\left(X^{2} + z^2 \right)^{-n -\frac{d}{2}}\label{T3expr}
\end{align}
The full kernel is, using \eqref{T1expr},\eqref{T2expr} and \eqref{T3expr},
\begin{align}
    \mathbf{K}_{n N} = z^{2 n}\left(T_{1} + T_{2} + T_{3}\right)
\end{align}
With this, we conclude this brief discussion on the mode kernels for the $p$- form field in empty AdS$_{d + 1}$ in arbitrary odd dimensions.
\section{Mode sum kernels in odd AdS$_{d + 1}$ : Graviton}\label{appC}

TThe odd AdS case for the graviton again has to be treated separately since for $d + 1 \in$ Odd, $\frac{d}{2}$ is an integer. Therefore the solution of the wave equation is given by the following expression
\begin{align}
    \Phi_{\mu \nu}  = \int_{|q| \geq 0}a_{\mu \nu}(q)z^{d/2}J_{d/2}(q z)e^{i q. x}\frac{\mathrm{d}^{d}q}{(2\pi)^{d}} + \int_{|q| \geq 0}b_{\mu \nu}(q)z^{d/2}Y_{d/2}(q z)e^{i q. x}\frac{\mathrm{d}^{d}q}{(2\pi)^{d}}\label{GravOdd}
\end{align}
We turn our attention to the odd AdS treatment for the $p$- form, which basically carries over to this case as well. This is a happy coincidence since the solutions \eqref{GravOdd} and \eqref{OddAdSsoln} are identical once one makes the identification $\nu = \frac{d}{2}$. 

For the sake of brevity, we write down the salient relations here. The power series expansion takes the form (where we use $2 n = d$, in order to be consistent with the notation)
\begin{align}
    \Phi_{\mu \nu}(z, x) = \sum_{k = 0}^{\infty}z^{2 n + 2 k}\phi_{k}(x) + \sum_{k = 0}^{\infty}\ln(z) z^{2 n + 2 k}\tilde{\phi}_{k}(x) + \sum_{k = 0}^{n - 1}z^{2 k}j_{k}(x)
\end{align}
where
\begin{align}
    j_{k}(x) &= \int_{q \geq 0}b_{\mu \nu}\beta_{k, n}2^{n}q^{2 k - n}e^{i q. x}\frac{\mathrm{d}^{d}q}{(2\pi)^{d}} \\
    \tilde{\phi}_{k}(x) &= \int_{q \geq 0}\frac{2}{\pi}b_{\mu \nu}\frac{(-1)^{k}q^{n + 2 k}}{2^{n}4^{k}\Gamma(k + 1)\Gamma(k + n + 1)}e^{i q. x}\frac{\mathrm{d}^{d}q}{(2\pi)^{d}} \\
    \phi_{k}(x) &= \int_{q \geq 0}a_{\mu \nu}\frac{(-1)^{k}q^{n + 2 k}}{2^{n}4^{k}\Gamma(k + 1)\Gamma(k + n + 1)}e^{i q. x}\frac{\mathrm{d}^{d}q}{(2\pi)^{d}} \notag \\ &+ \int_{q \geq 0}b_{\mu \nu}\left(\frac{2}{\pi}\ln(\frac{q}{2})\frac{(-1)^{k}q^{n + 2 k}}{2^{n}4^{k}\Gamma(k + 1)\Gamma(k + n + 1)} - \frac{\alpha_{k, n}}{2^{n}}q^{n + 2 k}\right)e^{i q. x}\frac{\mathrm{d}^{d}q}{(2\pi)^{d}}
\end{align}
where $\alpha_{k, n} = \frac{(-1)^{k}}{4^{k}\Gamma(k + 1)\Gamma(k + n + 1)}(\psi(k + 1) + \psi(k + n + 1))$ and $\beta_{k, n} = \frac{\Gamma(n - k)}{\Gamma( k + 1)4^{k}}$. 

Note that the only major (non-technical) difference with the $p$- form case is that the nature of the boundary data is different. So long as we are careful of the boundary data, we can adapt the expressions derived in the previous sections. The two kernels can then be read-off as before
\begin{align}
    \mathbf{K}_{N}(z,x; x') &= z^{n}2^{n}\Gamma(n + 1)\int_{q \geq 0}q^{-n}J_{n}(q z)e^{i q.(x - x')}\frac{\mathrm{d}^{d}q}{(2\pi)^{d}} \label{NormOddAdSkernelGrav} \\
    \mathbf{K}_{n N}(z, x; x') &= \frac{z^{n}}{2^{n}\Gamma(n)}\int_{q \geq 0}q^{n}Y_{n}(q z)e^{i q. (x - x')}\frac{\mathrm{d}^{d}q}{(2\pi)^{d}}\notag\\ &- \frac{z^{n}}{2^{n}\Gamma(n)}\int_{q \geq 0} \left( \frac{2 \ln(q/2)}{\pi}  + \gamma - \psi(n + 1)\right)q^{n}J_{n}(q z)e^{i q. (x - x')}\frac{\mathrm{d}^{d}q}{(2\pi)^{d}} \label{nonNormOddAdSkernelGrav}
\end{align}
These integrals can be similarly evaluated as before to give us the following expression
\begin{align}
    U_{1} &= -\frac{(-1)^{d/2}z^{-2 n - d}n\Gamma(d/2)}{ 2 \pi^{\frac{d}{2} + 2}(n + \frac{d}{2})}\;_{2}F_{1}\left(n + \frac{d}{2}, \frac{d}{2}; 1 + n + \frac{d}{2}; 1 + \frac{X^{2}}{z^2}\right)\notag\\
    &\xrightarrow[n = \frac{d}{2}]{} -\frac{(-1)^{d/2}d\Gamma(d/2)}{ 4 \pi^{\frac{d}{2} + 2}d}\;_{2}F_{1}\left(d, \frac{d}{2}; 1 + d; 1 + \frac{X^{2}}{z^2}\right)\label{U1expr}
\end{align}
The next term is
\begin{align}
    U_{2}
    &= -\frac{\Gamma \left(\frac{d}{2}+n\right)}{2 \pi^{\frac{d}{2} + 2}\Gamma(n)} \left(z^{2} + X^{2} \right)^{-\frac{d}{2}-n} \left(\psi\left(\frac{d}{2}+n\right)+\psi(n+1)-2 \log (X)+\log (4)\right) \notag \\
    &- \frac{ X^{-d - 2 n-1} \Gamma \left(\frac{d}{2}+n\right)}{2 \pi^{\frac{d}{2} + 2}\Gamma(n)} \left(G^{1}\left(n+1,\frac{d}{2}+n,n+1,-\frac{z^2}{X^2}\right)+G^{2}\left(n+1,\frac{d}{2}+n,n+1,-\frac{z^2}{X^2}\right)\right) \notag\\
    &\xrightarrow[n = \frac{d}{2}]{} \frac{\Gamma \left(d\right)}{2 \pi^{\frac{d}{2} + 2}\Gamma(\frac{d}{2})} \left(z^{2} + X^{2} \right)^{-d} \left(\psi\left(d\right)+\psi(\frac{d}{2}+1)-2 \log (X)+\log (4)\right) \notag \\
    &- \frac{ X^{-2 d -1} \Gamma \left(d\right)}{2 \pi^{\frac{d}{2} + 2}\Gamma(\frac{d}{2})} \left(G^{1}\left(\frac{d}{2}+1,d,\frac{d}{2}+1,-\frac{z^2}{X^2}\right)+G^{2}\left(\frac{d}{2}+1,d,\frac{d}{2}+1,-\frac{z^2}{X^2}\right)\right)\label{U2expr}
\end{align}
and the final term is 
\begin{align}
    U_{3} &= \left(\psi(n + 1) + \frac{2 \ln 2}{\pi} - \gamma \right)\frac{\Gamma(n + \frac{d}{2})}{2 \pi^{\frac{d}{2} + 1}\Gamma(n)}\left(X^{2} + z^2 \right)^{-n -\frac{d}{2}}\notag\\
    &\xrightarrow[n = \frac{d}{2}]{}\left(\psi(\frac{d}{2} + 1) + \frac{2 \ln 2}{\pi} - \gamma \right)\frac{\Gamma(d)}{2 \pi^{\frac{d}{2} + 1}\Gamma(\frac{d}{2})}\left(X^{2} + z^2 \right)^{-d}\label{U3expr}
\end{align}
And the full kernel for the non-normalizable mode is given by
\begin{align}
    \mathbf{K}_{nN} (z, x; x') = z^{d}(U_{1} + U_{2} + U_{3})
\end{align}
The kernel corresponding to the normalizable mode is the same as the even AdS case, now given by
\begin{align}
    \mathbf{K}_{N} = \frac{\Gamma(\frac{d}{2})}{2 \pi^{d/2 + 1}}\frac{z^{d}}{X^{d}}\;_{2}F_{1}\left(\frac{d}{2}, 1; \frac{d}{2} + 1; -\frac{z^2}{X^2}\right)
\end{align}
With this, we conclude this brief discussion on the mode kernels for the graviton in empty AdS$_{d + 1}$ in arbitrary odd dimensions.
\bibliographystyle{JHEP}
\bibliography{reference.bib}
\end{document}